\documentclass[journal]{IEEEtran}
\usepackage[noadjust]{cite}
\usepackage{graphicx}
\usepackage{amsthm}
\usepackage{amsmath}
\usepackage{tcolorbox}
\usepackage[ruled,vlined,linesnumbered]{algorithm2e}
\usepackage{algpseudocode}
\usepackage{amssymb}
\usepackage{tabularx}
\usepackage{url}
\usepackage{lipsum}

\usepackage{array}
\newcommand{\PreserveBackslash}[1]{\let\temp=\\#1\let\\=\temp}
\newcolumntype{C}[1]{>{\PreserveBackslash\centering}p{#1}}
\newcolumntype{R}[1]{>{\PreserveBackslash\raggedleft}p{#1}}
\newcolumntype{L}[1]{>{\PreserveBackslash\raggedright}p{#1}}

\usepackage{color,soul}
\soulregister\cite7
\soulregister\ref7
\soulregister\pageref7

\newcolumntype{P}[1]{>{\centering\arraybackslash}p{#1}}
\newcolumntype{M}[1]{>{\centering\arraybackslash}m{#1}}
\ifCLASSINFOpdf
\else
\fi
\SetArgSty{textnormal}

\begin{document}

\title{CPAR: Cloud-Assisted Privacy-preserving Image Annotation with Randomized KD-Forest}
\author{Yifan~Tian,~\IEEEmembership{Student~Member,~IEEE,}
        Yantian~Hou,~\IEEEmembership{Member,~IEEE,}
        and~Jiawei~Yuan,~\IEEEmembership{Member,~IEEE}    
\IEEEcompsocitemizethanks{\IEEEcompsocthanksitem The preliminary version of this paper appeared in the 5th IEEE Conference on Communications and Network Security (IEEE CNS 2017) \cite{tian2017capia}.

Yifan Tian and Jiawei Yuan are with the Department of ECSSE, Embry-Riddle Aeronautical University, e-mail: tiany1@my.erau.edu, yuanj@erau.edu. Yantian Hou is with the Computer Science Department, Boise State University, e-mail: yantianhou@boisestate.edu}}

\maketitle

\begin{abstract}
With the explosive growth in the number of pictures taken by smartphones, organizing and searching pictures has become important tasks. To efficiently fulfill these tasks, the key enabler is annotating images with proper keywords, with which keyword-based searching and organizing become available for images. Currently, smartphones usually synchronize photo albums with cloud storage platforms, and have their images annotated with the help of cloud computing. However, the ``offloading-to-cloud'' solution may cause privacy breach, since photos from smart photos contain various sensitive information. For privacy protection, existing research made effort to support cloud-based image annotation on encrypted images by utilizing cryptographic primitives. Nevertheless, for each annotation, it requires the cloud to perform linear checking on the large-scale encrypted dataset with high computational cost. 

This paper proposes a \underline{c}loud-assisted \underline{p}rivacy-preserving image \underline{a}nnotation with \underline{r}andomized kd-forest, namely CPAR. With CPAR, users are able to automatically assign keywords to their images by leveraging the power of cloud with privacy protected. CPAR proposes a novel privacy-preserving randomized kd-forest structure, which significantly improves the annotation performance compared with existing research. Thorough analysis is carried out to demonstrate the security of CPAR. Experimental evaluation on the well-known IAPR TC-12 dataset validates the efficiency and effectiveness of CPAR.
\end{abstract}


\IEEEpeerreviewmaketitle

\section{Introduction}
\IEEEPARstart{T}{he} widespread use of smartphones causes photography boom in recent years. According to a recent report from Forever's Strategy \& Business Development team \cite{forever2018}, the number of photos taken by smartphone is estimated to be 8.8 trillion in 2018. To facilitate the storage of photos, majority of smartphones today are synchronizing their photo albums with cloud storage, such as Apple's iCloud, Samsung Cloud, and Google Photos. Besides the storage service, these cloud storage platforms also help annotate users' photos with proper keywords, which is the key enabler for users to perform popular keyword-based search and organization over their photos. Although the cloud storage offers a set of decent features, it also raises privacy concerns since many users' photos may contain sensitive information, such as personal identities, locations, and financial information \cite{cloudprivacy}. To protect the privacy of photos, encrypting them with standard encryption algorithms, e.g., AES, is still the major approach for privacy protection in cloud storage \cite{Boxcryptor,S3-Encryption}. However, this kind of encryption also sacrifices many other attractive functionalities of cloud storage, especially for keyword-based search and management for imagery files.

In order to enable keyword-based search and management on encrypted data in cloud, keyword-based searchable encryption (SE) has been widely investigated in recent years \cite{Song:2000,EUROCRYPT04,WangCong:2010,Sun-2013AsiaCCS,Wang-2014Infocom}. An SE scheme typically provides encrypted search indexes constructed based on proper keywords assigned to data files. With these encrypted indexes, the data owner can submit encrypted keyword-based search request to search their data over ciphertexts. Unfortunately, these SE schemes all assume that keywords are already available for files to be processed, which is hard to be true for photos taken by smartphones. Specifically, unlike text files that support automatic keyword extraction from their contents, keywords assignment for imagery files relies on manual description or automatic annotation based on a large-scale pre-annotated image dataset. From the perspective of user experience, manually annotating each image from users' devices is clearly an impractical choice. Meanwhile, automatic image annotation that involves large-scale image datasets is too resource-consuming to be developed on smartphones. Although currently several cloud storage platforms offer image annotation services \cite{googlevision,scaleapi}, these platforms require access to unencrypted images. Therefore, how to provide efficient and privacy-preserving automatic annotation for smartphones' photos becomes the foundation of SE schemes applications on smartphones. To address this problem, our preliminary research proposes a scheme called CAPIA \cite{tian2017capia}. By tailoring homomorphic encryption over vector space, CAPIA offloads the image annotation process to the public cloud in privacy-preserving manner. Nevertheless, for every single annotation request, CAPIA requires the linear processing of all encrypted records in a large-scale dataset, which hence becomes its performance bottleneck for practical usage. 

This paper proposes a cloud-based privacy-preserving image annotation scheme using the power of cloud computing with significantly enhanced efficiency, namely CPAR. To turbocharge the annotation efficiency with privacy protected, CPAR designs a novel privacy-preserving randomized kd-forest structure. Specifically, CPAR first integrates operations for image annotation with the data search using randomized kd-forest \cite{muja2014scalable}. Then, by proposing a set of privacy-preserving comparison schemes, CPAR enables the cloud server to perform image annotation directly over an encrypted randomized kd-forest structure. Compared with the existing solution - CAPIA, CPAR offers an adjustable speedup rate from $4\times$ to $43.1\times$ while achieving $97.7\%$ to $80.3\%$ accuracy of CAPIA. Our privacy-preserving randomized kd-forest design can also be used as independent tools for other related fields, especially for these requiring similarity measurement on encrypted data. Moreover, considering the same keyword may have different importance for the semantic description of different images, CPAR also proposes a privacy-preserving design for real-time keywords ranking. To evaluate CPAR, thorough security analysis and numerical analysis are carried out first. Then, we implement a prototype of CPAR and conduct an extensive experimental evaluation using the well-known IAPR TC-12 dataset \cite{IAPR}. Our evaluation results demonstrate the practical performance of CPAR in terms of efficiency and accuracy.





The rest of this paper is organized as follows: In Section \ref{s:model}: we present the system model and threat model of CPAR. Section \ref{s:preliminaries} introduces backgrounds of automatic image annotation and technical preliminaries for CPAR. The detailed construction of CPAR is provided in Section \ref{s:detailedconstruction}. We analyze the security of CPAR in Section \ref{s:securityanalysis}. Section \ref{s:evaluation} evaluates the performance of CPAR. We review and discuss related works in Section \ref{s:related-work} and conclude this paper in Section \ref{s:conclusion}.

\section{Models}\label{s:model}
\subsection{System Model}\label{ss:systemmodel}
\begin{figure}[!ht]
\begin{center}
\includegraphics[height=3.7cm]{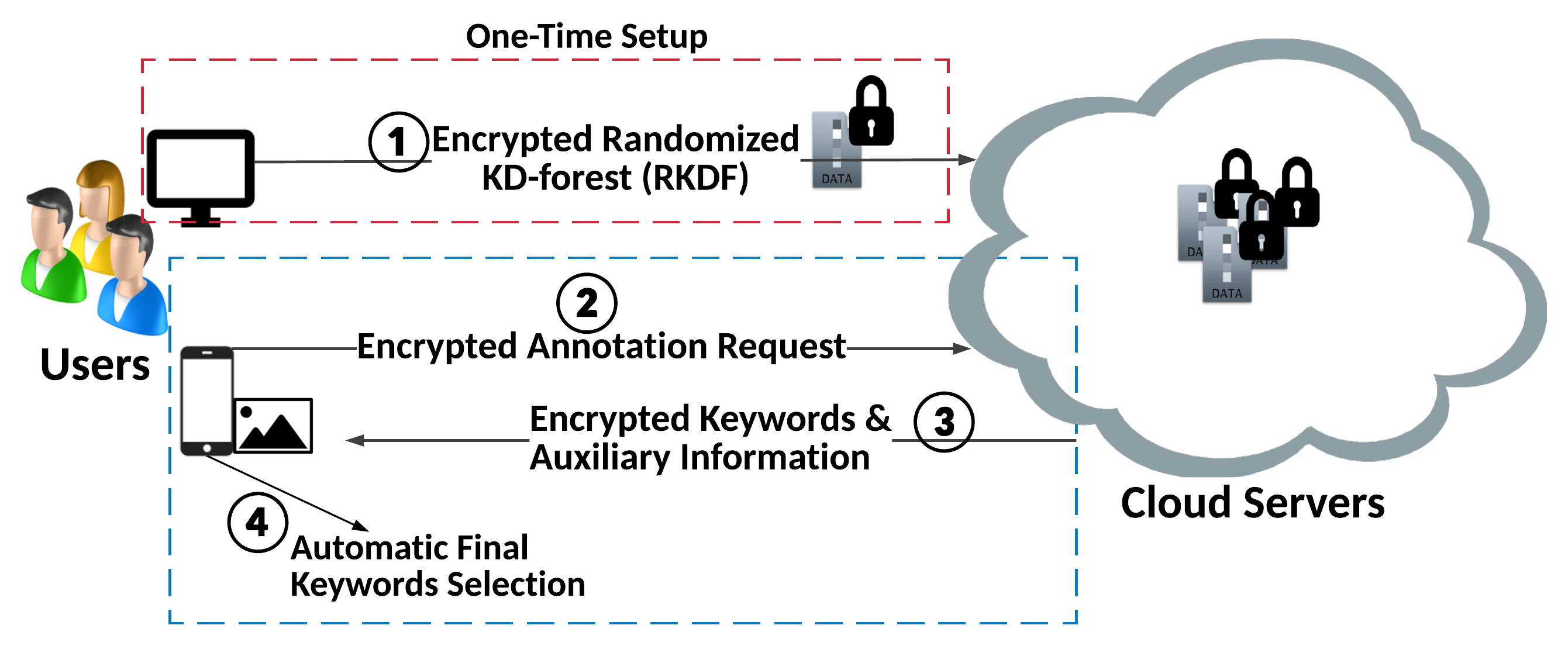}
\end{center}
\vspace{-5mm}
\caption{System Model of CPAR}\label{f:system-model}
\vspace{-5mm}
\end{figure}

As shown in Fig.\ref{f:system-model}, CAPR is composed of two entities: a \textit{Cloud Server} and a \textit{User}. The user stores his/her images on cloud, and the cloud helps the user to annotate his/her images without learning the contents and keywords of images. In CPAR, the user first performs a one-time system setup that constructs an encrypted randomized kd-forest with a pre-annotated image datasets. This encrypted randomized kd-forest is offloaded to the cloud server to assist future privacy-preserving image annotation. For resource-constrained mobile devices, this one-time setup process can be performed using desktops. Later on, when the user has a new image to annotate, he/she generates an encrypted request and sends it to the cloud. After processing the encrypted request, the cloud returns ciphertexts of top related keywords and auxiliary information to the user. Finally, the user decrypts all keywords and ranks them based on their real-time weights to select final keywords.

\subsection{Threat Model}\label{ss:treat-model}
In CPAR, we consider the cloud server to be ``curious-but-honest'', i.e., the cloud server will follow our scheme to perform storage and annotation services correctly, but it may try to learn sensitive information in user's data. The cloud server has access to all encrypted images, encrypted image features, encrypted keywords, encrypted RKDF, the user's encrypted requests, and encrypted annotation results. We also assume the user's devices are fully trusted and will not be compromised. The research on protecting user devices is orthogonal to this work. These assumptions are consistent with major research works that focus on search over encrypted data on public cloud \cite{WangCong:2010,Sun-2013AsiaCCS,Wang-2014Infocom}. CPAR focuses on preventing the cloud server from learning following information: 1) contents of the user's images; 2) features extracted and keywords annotated for each image; 3) request linkability, i.e., tell whether multiple annotation requests are from the same image.

\section{Preliminaries}\label{s:preliminaries}
\subsection{Image Feature Extraction}\label{ss:featureExt}
In this paper, we adopt global low-level image features as that are utilized in the baseline image annotation technique \cite{Makadia2010IJCV}, because it can be applied to general images without complex models and subsequent training. Color features of an image are extracted in three different color spaces: RGB, HSV, and LAB. In particular, RGB feature is computed as a normalized 3D histogram of RGB pixel, in which each channel (R,G,B) has 16 bins that divide the color space values from 0 to 255. The HSV and LAB features can be processed similarly as RGB, and thus we can construct three feature vectors for RGB, HSV and LAB respectively as $\textbf{V}_{RGB}$, $\textbf{V}_{HSV}$, and $\textbf{V}_{LAB}$. Texture features of an image are extracted using Gabor and Haar wavelets. Specifically, an image is first filtered with Gabor wavelets at three scales and four orientations, resulting in twelve response images. Each response image is then divided into non-overlapping rectangle blocks. Finally, mean filter response magnitudes from each block over all response images are concatenated into a feature vector, denoted as $\textbf{V}_{G}$. Meanwhile, a quantized Gabor feature of an image is generated using the mean Gabor response phase angle in non-overlapping blocks in each response image. These quantized values are concatenated into a feature vector, denoted as $\textbf{V}_{GQ}$. The Haar feature of an image is extracted similarly as Gabor, but based on differently configured Haar wavelets. HaarQ stands for the quantized version of Haar feature, which quantizes Haar features into [0,-1,1] if the signs of Haar response values are zero, negative, and positive respectively. We denote feature vectors of Haar and HaarQ as $\textbf{V}_H$ and $\textbf{V}_{HQ}$ respectively. Therefore, given an image, seven feature vectors will be extracted as $[\textbf{V}_{RGB},\textbf{V}_{HSV},\textbf{V}_{LAB},\textbf{V}_{G},\textbf{V}_{GQ},\textbf{V}_{H},\textbf{V}_{HQ}]$. For more details about the adopted image feature extraction, please refer to ref \cite{Makadia2010IJCV}.

\subsection{Integer Vector Encryption (IVE)}\label{ss:ive}
In this section, we describe a homomorphic encryption scheme designed for integer vectors \cite{zhou2014efficient}, which will be tailored in our construction to achieve privacy-preserving image annotation. For expression simplicity, following definitions will be used in the rest of this paper:
\begin{itemize}
  \item For a vector $\textbf{V}$ (or a matrix $\textbf{M}$), define $|max(\textbf{V})|$ (or $|max(\textbf{M})|$) to be the maximum absolute value of its elements.
  \item For $a \in \mathbb{R}$, define ${\lceil a \rfloor}$ to be the nearest integer of $a$, ${\lceil a \rfloor}_q$ to be the nearest integer of $a$ with modulus $q$.
  \item For matrix $\textbf{M} \in \mathbb{R}^{n \times m}$, define $vec(\textbf{M})$ to be a $nm$-dimensional vector by concatenating the transpose of each column of $\textbf{M}$.
\end{itemize}

\noindent \textbf{Encryption:} Given a $m$-dimensional vector $\textbf{V}\in \mathbb{Z}_p^{m}$ and the secret key matrix $\textbf{S}\in \mathbb{Z}^{m \times m}_q$, output the ciphertext of $\textbf{V}$ as
\begin{eqnarray}\label{e:ive-enc}
\textbf{C}(\textbf{V}) = \textbf{S}^{-1}(w\textbf{V}+\textbf{e})^T
\end{eqnarray}
where $\textbf{S}^{-1}$ is the inverse matrix of $\textbf{S}$, $T$ is the transpose operator, $\textbf{e}$ is a random error vector, $w$ is an integer parameter, $q>>p$, $w > 2|max(\textbf{e})|$.

\noindent \textbf{Decryption:} Given the ciphertext $\textbf{C}(\textbf{V})$, it can be decrypted using $\textbf{S}$ and $w$ as $\textbf{V} = \lceil \frac{(\textbf{S}\textbf{C}(\textbf{V}))^T}{w} \rfloor_q$.

\noindent \textbf{Inner Product:} Given two ciphertexts $\textbf{C}(\textbf{V}_1), \textbf{C}(\textbf{V}_2)$ of $\textbf{V}_1, \textbf{V}_2$, and their corresponding secret keys $\textbf{S}_{1}$ and $\textbf{S}_{2}$, the inner product operation of $\textbf{V}_1$ and $\textbf{V}_2$ over ciphertexts can be performed as 
\begin{equation}\label{ive-decryption} 
vec(\textbf{S}_{1}^{T}\textbf{S}_{2}) \lceil \frac{vec(\textbf{C}(\textbf{V}_1)\textbf{C}(\textbf{V}_2)^{T})}{w} \rfloor_q = w\textbf{V}_1\textbf{V}_2^T + \textbf{e}
\end{equation}
To this end, $vec(\textbf{S}_{1}^{T}\textbf{S}_{2})$ becomes the new secret key and $\lceil \frac{vec(\textbf{C}(\textbf{V}_1)\textbf{C}(\textbf{V}_2)^{T})}{w} \rfloor_q$ becomes the new ciphertext of $\textbf{V}_1\textbf{V}_2^T$. 

More details about this IVE encryption algorithm and its security proof are available in ref \cite{zhou2014efficient}.



\subsection{Order-Preserving Encryption (OPE)}\label{ss:opes}
Order-preserving symmetric encryption (OPE) is a deterministic encryption scheme whose encryption function preserves numerical ordering of the plaintexts. Given two integers $a$ and $b$ in which $a < b$, by encrypting with OPE, the order of $a$ and $b$ is preserved as $OPE(a) < OPE(b)$. More details about this OPE encryption scheme and its security proof are available in ref \cite{OPE,OPE3}.

\section{Construction of CPAR}\label{s:detailedconstruction} 
\subsection{Scheme Overview}\label{ss:overview}
The core idea of automatic image annotation is built on the hypothesis that images contain similar objects are likely to share keywords. The distance between the feature vectors of two images is used to measure the probability that they contain similar objects \cite{Makadia2010IJCV}. Given a large-scale pre-annotated image dataset, the annotation process for a new image can be treated as a process of finding a set of images with shared objects and transferring keywords from those images. As a result, the annotation efficiency becomes heavily dependent on the performance of finding the image with shared objects. To boost the search efficiency, CPAR adopts randomized kd-forest as the searching index \cite{muja2014scalable}. In addition, novel privacy-preserving schemes are designed to address the privacy concerns when integrating the randomized kd-forest into CPAR. Different from many other index structures that are only efficient for low-dimensional data, Randomized kd-forest (RKDF) is featured by its performance in handling high-dimensional data. In CPAR, data vectors are over 1300-dimension and thus making RKDF an effective selection.

\begin{figure*}[!th]
\begin{center}
\includegraphics[height=7cm]{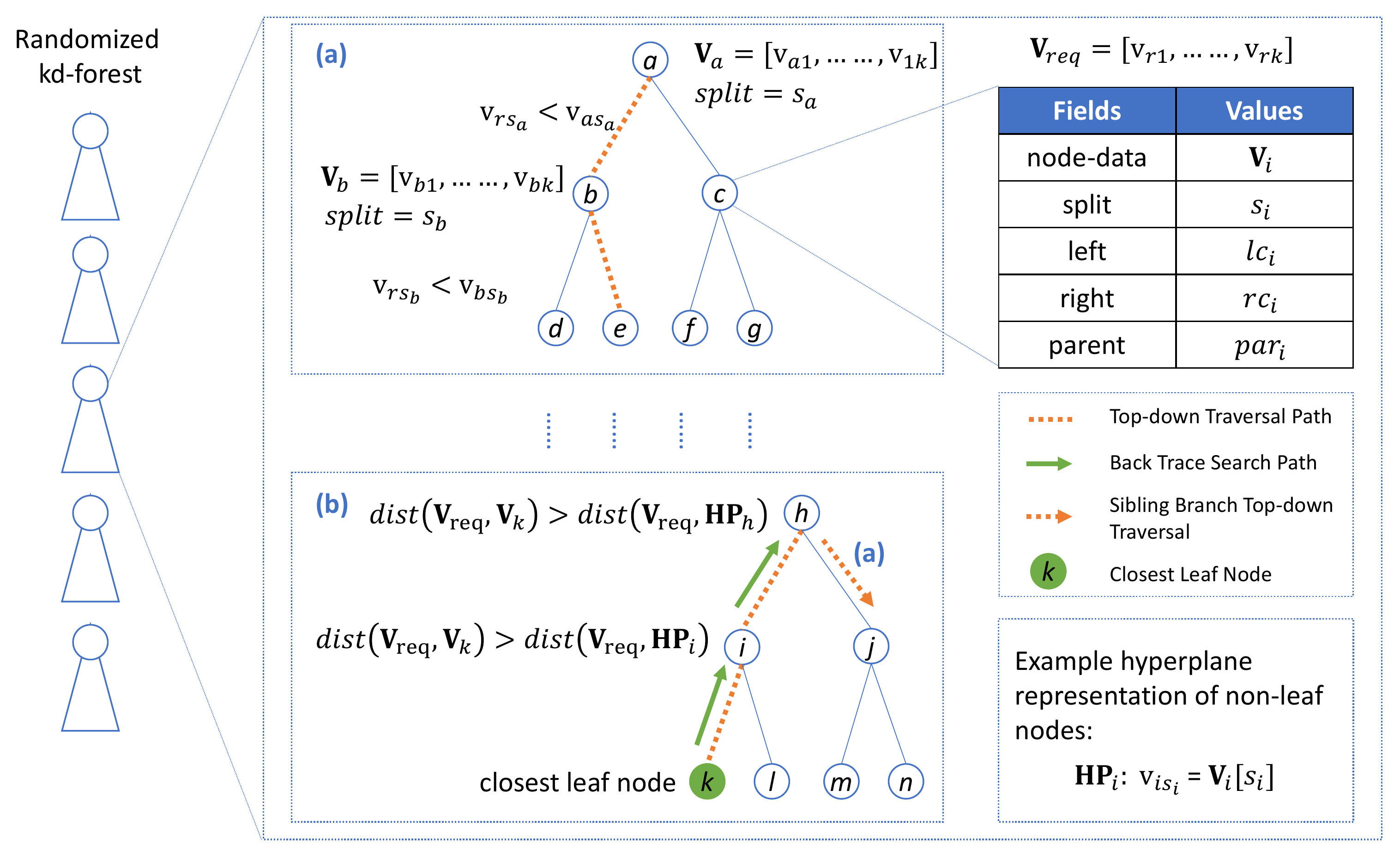}
\end{center}
\vspace{-5mm}
\caption{$\textbf{V}_{req}$ is the request vector and each $\textbf{V}_{i}$ is stored in each tree node $i$. $Dis(\cdot)$ is an arbitrary distance calculation function and $Dis(\textbf{V}_{req}, \textbf{H}_{i})$ is the distance between the request vector $\textbf{V}_{req}$ and $\textit{Node}_{i}$'s hyperplane. $\textbf{V}_{qL}$ is the least closest vector to $\textbf{V}_{req}$ in priority queue $Queue$. (a) represents top-down traversal; (b) represents back trace search and (c) represents queue push process.}\label{f:kd-forest} 
\vspace{-5mm}
\end{figure*}

As depicted in Figure \ref{f:kd-forest}, a RKDF is composed of a set of parallel kd-trees. For each $\textit{Node}_{i}$ in a kd-tree \cite{bentley1975multidimensional}, it stores a feature vector $\textbf{V}_{i}$ of dataset image $\textit{I}_{i}$. In addition, each non-leaf node also stores a $\textit{split}$ field $\textit{s}_{i}$ to generate a hyperplane that divides the vector space into two parts. Each $\textit{Node}_{j}$ in left sub-tree of $\textit{Node}_{i}$ has $\textit{Node}_{j}[s_i] \leq \textit{Node}_{i}[s_i]$ and vice versa, as described in ref \cite{bentley1975multidimensional}. To search nodes that store vectors with top-smallest distances to a request vector $\textbf{V}_{req}$, a parallel search among all trees in the forest is performed. Specifically, each tree is traversed in a top-down manner by comparing the $\textit{split}$ field values of $\textbf{V}_{req}$ and the vector $\textbf{V}_{i}$ stored in each $\textit{Node}_{i}$ as an example shown in Fig.\ref{f:kd-forest}(a). The traversal selects the left branch to continue if $\textbf{V}_{req}[\textit{s}_{i}] \leq \textbf{V}_{i}[\textit{s}_{i}]$ and vice versa. Once the traversal reaches a leaf node, the vector stored in that leaf node is pushed into a priority queue $Queue$ as a current close candidate to $\textbf{V}_{req}$. The queue push process is shown in Fig.\ref{f:kd-forest}(c). Note that during the search process, this $Queue$ keeps updating to hold $L$ closest vectors to $\textbf{V}_{req}$ and is shared by all trees in the forest. After that, a back trace search starts by iterating all the nodes in the path from the parent of the current node to the root node as an example shown in Fig.\ref{f:kd-forest}(b). When reaching a $\textit{Node}_{i}$ during the back trace, a same queue push is executed to judge whether to add $\textit{Node}_{i}$ to $Queue$ as illustrated in Fig.\ref{f:kd-forest}(c). For each $\textit{Node}_{i}$ in this path, a distance comparison between $Dis(\textbf{V}_{req}, \textbf{H}_{i})$ and $Dis(\textbf{V}_{req}, \textbf{V}_{qL})$ is compared, where $Dis(\textbf{V}_{req}, \textbf{H}_{i})$ is the distance between $\textbf{V}_{req}$ and a $\textit{Node}_{i}$'s hyperplane. $\textbf{H}_{i}$ can be considered as the projection vector of $\textbf{V}_{req}$ on $\textit{Node}_{i}$'s hyperplane. $\textbf{V}_{qL}$ is the $L$th vector in $Queue$ which meets $Dis(\textbf{V}_{req}, \textbf{V}_{qi}) \leq Dis(\textbf{V}_{req}, \textbf{V}_{qL}), \forall \textbf{V}_{qi} \in Queue$. If $Dis(\textbf{V}_{req}, \textbf{H}_{i})>Dis(\textbf{V}_{req}, \textbf{V}_{qL})$, the back trace continues to the next node in this path. Otherwise, the sibling branch of $\textit{Node}_{i}$ needs to be searched using the top-down traversal. In RKDF, once a node has been searched in one kd-tree, it will be marked and does not need to be checked again in the other trees. To further enhance the search efficiency of a RKDF, approximated search strategy can be adopted. In particular, based on the hypothesis that feature vectors of similar images are likely to be grouped in the same branch, there is a high probability that the targeted optimal top similar vectors will be visited well before visiting all nodes in each kd-tree. In Section \ref{s:evaluation}, we will evaluate the relationship among the approximation strength, accuracy, and efficiency. The detailed search of a RKDF is provided in Algorithm \ref{a:kd-forest-search}. For more details about the RKDF, please refer to ref \cite{muja2014scalable}.

To protect the privacy of user's data during the cloud-based annotation, the image data associated with the RKDF need to be encrypted. Furthermore, these encrypted data shall support corresponding search operations in RKDF, which include:
\begin{itemize}
  \item The comparison between $\textbf{V}_{req}[\textit{s}_{i}]$ and $\textbf{V}_{i}[\textit{s}_{i}]$ in the top-down traversal for path selection.
  \item The comparison between $Dis(\textbf{V}_{req}, \textbf{H}_{i})$ and $Dis(\textbf{V}_{req}, \textbf{V}_{qL})$ during the back trace process.
  \item The comparison between $Dis(\textbf{V}_{req}, \textbf{V}_{a})$ and $Dis(\textbf{V}_{req}, \textbf{V}_{b})$, i.e., distances from the request vector to two different images' feature vectors, which is used in the queue push process. 
\end{itemize}
The distance $Dis(\cdot)$ between two vectors is calculated with a combination of $L_1$ distance and KL-Divergence \cite{Makadia2010IJCV}. Specifically, the distance $Dis_{ab}$ of two vectors is computed as 
\begin{equation} 
\begin{aligned}
Dis_{ab}=&DL1^{RGB}_{ab}+DL1^{HSV}_{ab}+DL1^G_{ab}+DL1^{GQ}_{ab}\\\nonumber
&+DL1^H_{ab}+DL1^{HQ}_{ab}+DKL^{LAB}_{ab}
\end{aligned}
\end{equation}
where each vector has seven low-level color and texture feature vectors as discussed in Section \ref{ss:featureExt}, and $DL1$ and $DKL$ denote $L_1$ distance and KL-Divergence of two vectors after data normalization.

In order to address the privacy challenges while utilizing RKDF for cloud-assisted automatic image annotation, a challenge needs to be resolved: The original privacy-preserving comparison scheme for $L_1$ distance ($PL1C$) and KL-Divergence ($PKLC$) in CAPIA cannot be simply re-used in CPAR. That's because $PL1C$ and $PKLC$ can only support the privacy-preserving distance comparison between two vectors. However, while searching in a RKDF, the distance comparison between a vector and a hyperplane needs to be supported in the back trace process and queue push process of RKDF. In order to resolve this issue, we re-design $PL1C$ and $PKLC$ to get $PL1C-RF$ and $PKLC-RF$, standing for $PL1C$ and $PKLC$ for RKDF. $PL1C-RF$ and $PKLC-RF$ enable the aforementioned privacy-preserving distance comparison between two vectors as well as between one vector and one hyperplane. In addition, we integrate order-preserving encryption \cite{OPE,OPE3} into CPAR to protect the comparison of $\textit{split}$ field values in the top-down traversal of RKDF.



\subsection{PL1C-RF: Privacy-preserving $L_1$ Distance Comparison for Randomized KD-Forest}\label{ss:PL1C-RF}
In $PL1C-RF$, we consider two types of $L_1$ distance comparison that are required in the queue push and back trace process of RKDF: 1) $DL1_{ac}$ and $DL1_{bc}$ for three image feature vectors $\textbf{V}_i,i\in\{a,b,c\}$; 2) $DL1_{hc}$ and $DL1_{bc}$ for a hyperplane projected vector $\textbf{H}_a$ and two image feature vectors $\textbf{V}_b,\textbf{V}_c$. $DL1_{hc}$ is measured by the $L_1$ distance between $\textbf{H}_a[s_a]$ and $\textbf{V}_c[s_a]$, where $s_a$ is the $\textit{split}$ field of the $\textit{Node}_{a}$. To be more specific, $DL1_{hc}$ is calculated by projecting $\textbf{V}_c$ on $\textit{Node}_a$'s hyperplane and then calculating the $L_1$ distance between $\textbf{V}_c$ and the projected vector $\textbf{H}_a$.


\textbf{Data Preparation}: Given an image feature vector $\textbf{V}_i=[v_{i1},\cdots,v_{im}]$, the user first converts it to a $m\beta$-dimensional binary vector $\tilde{\textbf{V}}_i=[F(v_{i1}),\cdots,F(v_{im})]$, where $\beta = |qL(\textbf{V}_i)|$, and $F(v_{ij})=[1,1,\cdots, 1,0,\cdots,0]$ such that the first $v_{ij}$ terms are 1 and the rest $\beta-v_{ij}$ terms are 0. The $L_1$ distance between $\textbf{V}_a$ and $\textbf{V}_b$ now can be calculated as 
\begin{center}
$DL1_{ab}=\sum_{j=1}^m|v_{aj}-v_{bj}|=\sum_{j=1}^{m\beta}(\tilde{v}_{aj}-\tilde{v}_{bj})^2$
\end{center}
Then, the approximation introduced in ref \cite{sec-l1} is applied to $\tilde{\textbf{V}}_i$ to update its dimension from $m\beta$ to $\hat{m}=\alpha m\log_\gamma^{\beta+1}$ based on the Johnson Lindenstrauss (JL) Lemma \cite{JL-Lemma}. By denoting the approximated vector as $\hat{\textbf{V}}_i$, we have $DL1_{ab}=\sum_{j=1}^{m\beta}(\tilde{v}_{aj}-\tilde{v}_{bj})^2 \approx \sum_{j=1}^{\hat{m}}(\hat{v}_{aj}-\hat{v}_{bj})^2$. The correctness and accuracy of such an approximation have been proved in ref \cite{sec-l1}. According to our experimental evaluation in Section \ref{s:evaluation}, we sets $\alpha = 1$ and $\gamma=100$ in CPAR to balance accuracy and efficiency.

The detailed construction of the rest stages in \textit{PL1C-RF} is presented in Fig.\ref{f:pl1c-Rf}. The user first encrypts the image feature vectors and its corresponding hyperplane projected vector (if exists), and then stores them in the cloud. Later on the user can generate encrypted $L_1$ distance comparison request and ask the cloud to conduct privacy-preserving comparison. On receiving the request, the cloud can conduct two types of $L_1$ distance comparison using ciphertext only according to user's request.

\begin{figure}[ht]
\begin{tcolorbox}[lowerbox=invisible,colback=white,boxrule=0.3mm]
\begin{center}
\textbf{Construction of PL1C-RF}
\end{center}
\footnotesize

\textbf{Data Encryption}: 
\begin{enumerate}
\item Append 3 elements to an approximated $\hat{\textbf{V}}_i$ as $\hat{\textbf{V}}_i=[\hat{v}_{i1},\hat{v}_{i2},\cdots, \hat{v}_{i\hat{m}},r-\frac{1}{2}\sum_{j=1}^{\hat{m}}{\hat{v}_{ij}}^2,\epsilon_i, -1], i \in \{a,b\}$, where $r$ is a random number and $\epsilon_i$ is a small random noise. 

\item If $\hat{\textbf{V}}_i$ is stored in a non-leaf node, generate a $(2\hat{m}+2)$-dimensional hyperplane projected vector as $\hat{\textbf{H}}_{i}=[0, \cdots, \hat{v}_{is_{i}}, \cdots, 0, r-\frac{1}{2}\hat{v}_{is_{i}}^2, \epsilon_{i}^{'}, 0, \cdots,0 -1,0 \cdots, 0]$, where $r-\frac{1}{2}\hat{v}_{is_{i}}^2$ is the $(\hat{m} + 1)$th element, $-1$ is the $(\hat{m} + 2 + s_{i})$th element, and $\textit{s}_{i}$ is the \textit{split} field of node $i$. 

\item Encrypt $\hat{\textbf{V}}_i$ and $\hat{\textbf{H}}_{i}$ using the \textbf{Encryption} algorithm of IVE as $\textbf{C}(\hat{\textbf{V}}_{i})=\textbf{S}^{-1}(w\hat{\textbf{V}}_{i}+\textbf{e}_i)^T$ and $\textbf{C}(\hat{\textbf{H}}_{i})=\textbf{S}^{'-1}(w\hat{\textbf{H}}_{i}+\textbf{e}_{i}^{'})^T$. $\textbf{C}(\hat{\textbf{V}}_{i}), \textbf{C}(\hat{\textbf{H}}_{i})$, and $w$ are outsourced to the cloud.
\end{enumerate}


\textbf{Request Generation}: 
\begin{enumerate}
\item Append approximated request vector $\hat{\textbf{V}}_c$ as $\hat{\textbf{V}}_c=[r_c\hat{v}_{c1},\cdots, r_c\hat{v}_{c\hat{m}},r_c, 1, \frac{1}{2}r_{c}\sum_{j=1}^{\hat{m}}{\hat{v}_{cj}}^2]$, in which $r_c$ is a positive random number. 

\item Generate $\hat{\textbf{H}}_{c}=[r_c\hat{v}_{c1},\cdots, r_c\hat{v}_{c\hat{m}},r_c, 1, \frac{1}{2}r_{c}\hat{v}_{c1}^{2}, \cdots, $ $ \frac{1}{2}r_{c}\hat{v}_{c\hat{m}}^{2}]$ as hyperplane projected vector. 

\item $\hat{\textbf{V}}_c$ and $\hat{\textbf{H}}_{c}$ are encrypted as $\textbf{C}(\hat{\textbf{V}}_c)=\textbf{S}_c^{-1}(w\hat{\textbf{V}}_c+\textbf{e}_c)^T$ and $\textbf{C}(\hat{\textbf{H}}_{c})=\textbf{S}_c^{'-1}(w\hat{\textbf{H}}_{c}+\textbf{e}_{c}^{'})^T$. $\textbf{C}(\hat{\textbf{V}}_c)$, $\textbf{C}(\hat{\textbf{H}}_{c})$, $\textbf{S}^{T}\textbf{S}_{c}$ and $\textbf{S}^{'T}\textbf{S}_{c}^{'}$ are sent to the cloud as request.
\end{enumerate}

\textbf{Distance Comparison}: \newline
\underline{Type-1: Compare $DL1_{ac}, DL1_{bc}$} 
\begin{enumerate}
\item Given $\textbf{C}(\hat{\textbf{V}}_a)$, $\textbf{C}(\hat{\textbf{V}}_b)$ and $\textbf{C}(\hat{\textbf{V}}_c)$, compute $\lceil \frac{vec(\textbf{C}(\hat{\textbf{V}}_a)\textbf{C}(\hat{\textbf{V}}_c)^T)}{w} \rfloor_q$, $ \lceil \frac{vec(\textbf{C}(\hat{\textbf{V}}_b)\textbf{C}(\hat{\textbf{V}}_c)^T)}{w} \rfloor_q$ and decrypt them as $\hat{\textbf{V}}_a\hat{\textbf{V}}_c^T$ and $\hat{\textbf{V}}_b\hat{\textbf{V}}_c^T$ as Eq.\ref{ive-decryption}. 
	
\item Compare the approximated $L_1$ distance comparison as $\hat{\textbf{V}}_b\hat{\textbf{V}}_c^T-\hat{\textbf{V}}_a\hat{\textbf{V}}_c^T \approx\frac{r_c}{2}(DL1_{ac}-DL1_{bc})+(\epsilon_b-\epsilon_a)$.
\end{enumerate}

\underline{Type-2: Compare $DL1_{hc}, DL1_{bc}$},  
\begin{enumerate}
\item Given $\textbf{C}(\hat{\textbf{H}}_{a})$, $\textbf{C}(\hat{\textbf{V}}_{b})$, $\textbf{C}(\hat{\textbf{V}}_{c})$ and $\textbf{C}(\hat{\textbf{H}}_{c})$, compute $\lceil \frac{vec(\textbf{C}(\hat{\textbf{H}}_{a})\textbf{C}(\hat{\textbf{H}}_{c})^{T})}{w} \rfloor_q$, $\lceil \frac{vec(\textbf{C}(\hat{\textbf{V}}_{b})\textbf{C}(\hat{\textbf{V}}_{c})^{T})}{w} \rfloor_q$ and decrypt them as $\hat{\textbf{H}}_a\hat{\textbf{H}}_{c}^{T}$ and $\hat{\textbf{V}}_{b}\hat{\textbf{V}}_{c}^{T}$ as Eq.\ref{ive-decryption}. 

\item Compare the approximated $L_1$ distance comparison as $ \hat{\textbf{V}}_{b}\hat{\textbf{V}}_{c}^{T}-\hat{\textbf{H}}_a\hat{\textbf{H}}_c^{T} \approx\frac{r_c}{2}(DL1_{hc}-DL1_{bc})+(\epsilon_b-\epsilon^{'}_a)$.
\end{enumerate}

\end{tcolorbox}
\caption{Construction of $PL1C-RF$}\label{f:pl1c-Rf}
\end{figure}

It is worth to note that $PL1C-RF$ is only interested in which distance is smaller during the comparison. Therefore, instead of letting the cloud get exact $L_1$ distances for comparison, $PL1C-RF$ adopts approximated distance comparison result scaled and obfuscated by $r_c$, $\epsilon_b-\epsilon_a$ and $\epsilon_b-\epsilon_a^{'}$ as shown in \ref{ss:PL1C-RF}. As $r_c$ is a positive random number, the sign of $\frac{r_c}{2}(DL1_{ac}-DL1_{bc})$ and $\frac{r_c}{2}(DL1_{hc}-DL1_{bc})$ are consistent with $DL1_{ac}-DL1_{bc}$ and $DL1_{hc}-DL1_{bc}$ respectively. Meanwhile, since $r_c>>\epsilon_b-\epsilon_a$ and $r_c>>\epsilon_b-\epsilon_a^{'}$, the added noise term has negligible influence to the sign of $DL1_{ac}-DL1_{bc}$ or $DL1_{hc}-DL1_{bc}$ unless these two distances are very close to each other. Fortunately, instead of finding the most related one, our CPAR design will utilize $PL1C-RF$ to figure out top 10 related candidates during the comparison. Such a design makes important candidates (say top 5 out of top 10) not be bypassed by the error introduced in $\epsilon_b-\epsilon_a$ and $\epsilon_b-\epsilon_a^{'}$. This hypothesis is further validated by our experimental results in Section \ref{s:evaluation}.

\subsection{PKLC-RF: Privacy-preserving KL-Divergence Comparison for Randomized KD-Forest}\label{ss:PKLC-RF}
In $PKLC-RF$, we also consider two types of KL-Divergence comparison similar to $PL1C-RF$: 1) $DKL_{ac}$ and $DKL_{bc}$ for three image feature vectors $\textbf{V}_i,i\in\{a,b,c\}$; 2) $DKL_{hc}$ and $DKL_{bc}$ for a hyperplane projected vector $\textbf{H}_a$ and two image feature vectors $\textbf{V}_b,\textbf{V}_c$. Given two $m$-dimensional vectors $\textbf{V}_i,i\in \{a,b\}$, their KL-Divergence $DKL_{ab}$ is calculated as
\begin{eqnarray}\label{e:plain-kl}
DKL_{ab}&=&\sum_{j=1}^m v_{aj} \times log(\frac{v_{aj}}{v_{bj}}) \\
&=&\sum_{j=1}^m v_{aj} \times log(v_{aj})-\sum_{j=1}^m v_{aj} \times log(v_{bj}) \nonumber
\end{eqnarray}
where $log(\frac{v_{aj}}{v_{bj}})=log(v_{aj})=log(v_{bj})=0$ if $v_{aj}=0$ or $v_{bj}=0$. In addition, the KL-Divergence $DKL_{hc}$ between a image feature vector and a hyperplane is measured by the KL-Divergence between $\textbf{H}_a[s_a]$ and $\textbf{V}_c[s_a]$, where $s_a$ is the \textit{split} field of $\textit{Node}_{a}$. Similar with $PL1C-RF$, $PKLC-RF$ is also calculated by projecting $\textbf{V}_c$ on $\textit{Node}_{a}$'s hyperplane and then calculating the KL-Divergence between $\textbf{V}_c$ and the projected vector $\textbf{H}_a$.

The detailed construction of $PKLC-RF$ is presented in Fig.\ref{f:pklc-Rf}. In the data encryption stage, the image feature vectors and corresponding hyperplane projected vector (if exists) are encrypted and stored in the cloud. On receiving the encrypted KL-Divergence comparison request from the user, the cloud conducts two types of privacy-preserving KL-Divergence comparison using ciphertext only according to user's request. Similar to our $PL1C$ construction, we have $r_c>0$ and $r_c>>(\epsilon_b-\epsilon_a)$. Therefore, the cloud can figure out which KL-Divergence is smaller based on the scaled and obfuscated comparison result.  

\begin{figure}[ht]
\begin{tcolorbox}[lowerbox=invisible,colback=white,boxrule=0.3mm]
\begin{center}
\textbf{Construction of PKLC-RF}
\end{center}
\footnotesize

\textbf{Data Encryption}: 
\begin{enumerate}
\item Given an image feature vector $\textbf{V}_i$, append $m+2$ elements as $\textbf{V}_i=[v_{i1},v_{i2},\cdots, v_{im},v_{i1}\times log(v_{i1}),\cdots,v_{im}\times log(v_{im}) ,r, \epsilon_i]$, where $r$ is a random number and $\epsilon_i$ is a small random noise. If $\textbf{V}_i$ is stored in a non-leaf node in RKDF, its corresponding hyperplane projected vector is processed as $\textbf{H}_{i}=[0, \cdots, v_{is_{i}}, \cdots, 0, \cdots, v_{is_{i}} \times log(v_{is_{i}}), \cdots, 0, r, \epsilon_i^{'}]$, where $s_{i}$ is the \textit{split} field of the node, $v_{is_{i}}$, $v_{is_{i}} \times log(v_{is_{i}})$ and $r$ are the $s_{i}$th, $(m+s_{i})$th and $(2m+1)th$ elements respectively.

\item Encrypt $\textbf{V}_i$ and $\textbf{H}_{i}$ with the \textbf{Encryption} algorithm of IVE as $\textbf{C}(\textbf{V}_i)=\textbf{S}^{-1}(w{\textbf{V}}_i+\textbf{e}_i)^T$ and $\textbf{C}(\textbf{H}_i)=\textbf{S}^{-1}(w{\textbf{H}}_{i}+\textbf{e}_{i}^{'})^T$. 
\end{enumerate}

\textbf{Request Generation}: 
\begin{enumerate}
\item Given request image feature vector $\textbf{V}_c$, replace its elements $v_{cj}$ with $-r_c\times log(v_{cj})$ and append $m+2$ elements to it as $\textbf{V}_c=[-r_c\times log(v_{c1}),\cdots, -r_c\times log(v_{cm}),G(v_{c1}),\cdots, G(v_{cm}), r_c,-1]$, where $G(v_{cj})=\left\{ \genfrac {}{} {0pt}{0}{r_c,v_{cj}\neq 0}{0,v_{cj}= 0}\right.$, $r_c$ is a positive random number changing for every request.

\item Using the \textbf{Encryption} algorithm of IVE to encrypt $\textbf{V}_c$ as $\textbf{C}(\textbf{V}_c)=\textbf{S}_c^{-1}(w{\textbf{V}}_c+\textbf{e}_c)^T$. $\textbf{C}(\textbf{V}_c)$ and $\textbf{S}^{T}\textbf{S}_{c}$ are sent to the cloud as request. 
\end{enumerate}

\textbf{KL-Divergence Comparison}: \newline
\underline{Type-1: Compare $DKL_{ac}, DKL_{bc}$}
\begin{enumerate}
\item Compute $ \lceil \frac{vec(\textbf{C}(\textbf{V}_a)\textbf{C}(\textbf{V}_c)^T)}{w} \rfloor_q$, $ \lceil \frac{vec(\textbf{C}(\textbf{V}_b)\textbf{C}(\textbf{V}_c)^T)}{w} \rfloor_q$ and decrypts them as $\textbf{V}_a\textbf{V}_c^T$ and $\textbf{V}_b\textbf{V}_c^T$ using the \textbf{Decryption} of IVE in Section \ref{ss:ive}.
\item Compare KL divergence as $\textbf{V}_a\textbf{V}_c^T-\textbf{V}_b\textbf{V}_c^T = r_c(DKL_{ac}-DKL_{bc})+(\epsilon_b-\epsilon_a)\nonumber$.
\end{enumerate}

\underline{Type-2: Compare $DKL_{hc}, DKL_{bc}$}
\begin{enumerate}
\item Compute $\lceil \frac{vec(\textbf{C}(\textbf{H}_{a})\textbf{C}(\textbf{V}_{c})^T)}{w} \rfloor_q$, $\lceil \frac{vec(\textbf{C}(\textbf{V}_{b})\textbf{C}(\textbf{V}_{c})^{T})}{w} \rfloor_q$ and decrypts as $\textbf{H}_a\textbf{V}_{c}^T$ and $\textbf{V}_{b}\textbf{V}_{c}^{T}$ using the \textbf{Decryption} of IVE as Eq.\ref{ive-decryption}.

\item Compare KL divergence as $\textbf{H}_a\textbf{V}_c^T-\textbf{V}_{b}\textbf{V}_c^T = r_c(DKL_{hc}-DKL_{bc})+(\epsilon_b-\epsilon_{a}^{'})\nonumber$.
\end{enumerate}
\end{tcolorbox}
\caption{Construction of PKLC-RF}\label{f:pklc-Rf}
\end{figure}

\subsection{Detailed Construction of CPAR}
CPAR consists of five major procedures. In the \textit{System Setup}, the user selects system parameters, extracts, pre-processes feature vectors of images in a pre-annotated dataset and uses these feature vectors to build a RKDF. Then, the user executes the \textit{RKDF Encryption} procedure to encrypt all data associated with nodes in the RKDF. Both the \textit{System Setup} procedure and the \textit{RKDF Encryption} procedure are one-time cost in CPAR. Later on, the user can use the \textit{Secure Annotation Request} procedure to generate an encrypted annotation request. On receiving the request, the cloud server performs the \textit{Privacy-preserving Annotation on Cloud} procedure to return encrypted keywords for the requested image. At the end, the user obtains final keywords by executing the \textit{Final Keyword Selection} procedure.

\subsubsection{System Setup}\label{s:systemsetup1}
To perform the one-time setup of CPAR system, the user first prepares a pre-annotated image dataset with $n$ images, which can be obtained from public sources, such as IAPR TC-12 \cite{IAPR}, LabelMe \cite{labelme}, etc. For each image $I_i$ in the dataset, the user extracts seven feature vectors $[\textbf{V}_{i,{RGB}},\textbf{V}_{i,{HSV}},\textbf{V}_{i,{LAB}},\textbf{V}_{i,{G}},\textbf{V}_{i,{GQ}},\textbf{V}_{i,{H}},\textbf{V}_{i,{HQ}}]$. Compared with other five feature vectors that have dimension up to 256, $\textbf{V}_{i,{H}}$ and $\textbf{V}_{i,{HQ}}$ have a high dimension as 4096. To guarantee the efficiency while processing feature vectors, Principal Component Analysis (PCA) \cite{PCA} is utilized to reduce the dimension of $\textbf{V}_{i,{H}}$ and $\textbf{V}_{i,{HQ}}$. According to our experimental evaluation in Section \ref{ss:real-time-anno}, PCA-based dimension reduction with proper setting can significantly improve the efficiency of CPAR with slight accuracy loss. After that, $L_1$ normalization will be performed for each feature vector, which normalizes elements in these vectors to [-1,1]. Besides $\textbf{V}_{i,{LAB}}$, the user also increases each element in $\textbf{V}_{i,k}, k\in \{RGB, HSV, G, GQ,H,HQ\}$ as $v_{i,k,j}=v_{i,k,j}+1$ to avoid negative values. Six feature vectors that use $L_1$ distance for similarity measurement are concatenated as a $m_{L1}$-dimensional vector $\textbf{V}_{i,L1}$. $\textbf{V}_{i, LAB}$ is denoted as a $m_{KL}$-dimensional vector $\textbf{V}_{i,KL}$ for expression simplicity. It is easy to verify that $DL1_{ab}^{L1}=DL1^{RGB}_{ab}+DL1^{HSV}_{ab}+DL1^G_{ab}+DL1^{GQ}_{ab}+DL1^H_{ab}+DL1^{HQ}_{ab}$. 

After that, a RKDF is constructed with feature vector space $\{\textbf{V}_i\}_{1 \leq i \leq n}$, in which each node in a single tree is associated with one $\textbf{V}_i$. For each non-leaf node in RKDF, its \textit{split} field element $\textbf{V}_i[s_i]$ is stored in a set $\mathcal{SF}$. In CPAR, the RKDF contains ten parallel kd-trees.

\subsubsection{RKDF Encryption}\label{s:dataenc1}
Given an image $I_i$ in the pre-annotated dataset, its keywords $\{K_{i,t}\}$ are first encrypted using AES. Then, its processed feature vectors $\textbf{V}_{i,L1}, \textbf{V}_{i,KL}$ are encrypted with our $PL1C-RF$ and $PKLC-RF$ schemes as $\textbf{C}(\textbf{V}_{i,L1})$ and $\textbf{C}(\textbf{V}_{i,KL})$ respectively. $\textbf{C}(\textbf{V}_{i,L1})$ and $\textbf{C}(\textbf{V}_{i,KL})$ are then stored in the corresponding $\textit{Node}_{i}$ of the RKDF. For each non-leaf node, encrypted hyperplane projected vectors $\textbf{C}(\textbf{H}_{i,L1}),\textbf{C}(\textbf{H}_{i,KL})$ are generated and added into $\textit{Node}_{i}$ using the data encryption processes described in our $PL1C-RF$ and $PKLC-RF$. In addition, for the \textit{split} field element $\textbf{V}_i[s_i]$ of each non-leaf node, an order-preserving encryption is executed and the ciphertext $OPE(\textbf{V}_i[s_i])$ is stored in $\textit{Node}_{i}$. After the encryption, each node in the RKDF only contains encrypted data as
\begin{itemize}
\small
\item \textbf{Non-leaf Node}: $[\textbf{C}(\textbf{V}_{i,L1}), \textbf{C}(\textbf{V}_{i,KL}),\textbf{C}(\textbf{H}_{i,L1}),\textbf{C}(\textbf{H}_{i,KL}),$ $OPE(\textbf{V}_i[s_i]),AES(\{K_{i,t}\})]$  
\item \textbf{Leaf Node}: $[\textbf{C}(\textbf{V}_{i,L1}), \textbf{C}(\textbf{V}_{i,KL}),AES(\{K_{i,t}\})]$
\end{itemize}
During the encryption process, same secret keys $\textbf{S}_{L1}$, $\textbf{S}_{L1}^{'}$, $\textbf{S}_{KL}$, public parameter $w$, and random number $r$ will be used for all images. However, different error vector $\textbf{e}_i, \textbf{e}_i^{'}$ and noise term $\epsilon_i, \epsilon_i^{'}$ are generated for each image $I_i$ correspondingly. The user also computes $\textbf{S}_{L1}^{T}\textbf{S}_{s,L1}$, $\textbf{S}_{L1}^{'T}\textbf{S}_{s,L1}^{'}$ and $\textbf{S}_{KL}^{T}\textbf{S}_{s,KL}$, in which $\textbf{S}_{s,L1}$, $\textbf{S}_{s,L1}^{'}$ and $\textbf{S}_{s,KL}$ are secret keys for the encryption of later annotation requests. The encrypted RKDF, $\textbf{S}_{L1}^{T}\textbf{S}_{s,L1}$, $\textbf{S}_{L1}^{'T}\textbf{S}_{s,L1}^{'}$ and $\textbf{S}_{KL}^{T}\textbf{S}_{s,KL}$ are outsourced to the cloud. 



\subsubsection{Secure Annotation Request}\label{s:securereq1}
When the user has a new image $I_s$ for annotation, he/she first extracts seven feature vectors as $\textbf{V}_s,s\in[RGB,HSV,LAB,G,GQ,H,HQ]$. These vectors will be processed to output $\textbf{V}_{s,L1}$ and $\textbf{V}_{s,KL}$ as that in the \textit{System Setup} procedure. $\textbf{V}_{s,L1}$ and $\textbf{V}_{s,KL}$ are encrypted as $\textbf{C}(\textbf{V}_{s,L1})$, $\textbf{C}(\textbf{H}_{s,L1})$, and $\textbf{C}(\textbf{V}_{s,KL})$ using the \textit{Request Generation} of $PL1C-RF$ and $PKLC-RF$ schemes respectively. For each annotation request, the user generates a new positive random number $r_s$ and new error vectors $\textbf{e}_s, \textbf{e}_s^{'}$. Meanwhile, for each element $sf_j$ in the \textit{split} field set $\mathcal{SF}$ generated in \textit{System Setup}, the user encrypts $\textbf{V}_{s}[sf_j]$ using order-preserving encryption as $OPE(\textbf{V}_{s}[sf_j])$. $\textbf{C}(\textbf{V}_{s,L1})$, $\textbf{C}(\textbf{H}_{s,L1})$, $\textbf{C}(\textbf{V}_{s,KL})$ and $\{OPE(\textbf{V}_{s}[sf_j])\}$ are sent to the cloud as the annotation request.


\subsubsection{Privacy-preserving Annotation on Cloud}\label{s:ppanotation1}
On receiving the encrypted request, the cloud first performs a privacy-preserving search over the encrypted RKDF. As described in Algorithm \ref{a:kd-forest-search}, the cloud conducts parallel search over each encrypted tree in the RKDF. There are three places that require the cloud to conduct privacy-preserving computation over encrypted data:

$\bullet$ During the top-down traversal, as the \textit{split} field element of each non-leaf node is encrypted using order-preserving encryption, the cloud can directly compare their ciphertexts (line 7) to determine which node to be checked next.

$\bullet$ In the back trace process, the cloud needs to perform privacy-preserving comparison to determine whether the current node's sibling branch needs to be searched (line 24 to 29). In particular, given $\textbf{C}(\textbf{V}_{s,L1})$, $\textbf{C}(\textbf{H}_{s,L1})$, $\textbf{C}(\textbf{V}_{qL,L1})$, $\textbf{C}(\textbf{H}_{parent,L1})$, $\textbf{C}(\textbf{V}_{s,KL})$, $\textbf{C}(\textbf{V}_{qL,KL})$, and $\textbf{C}(\textbf{H}_{parent,KL})$, the cloud first uses type-2 distance comparison in $PL1C-RF$ and $PKLC-RF$ to compute

\begin{center}
$\textbf{V}_{qL,L1}\textbf{V}_{s,L1}^T, ~~~\textbf{V}_{qL,KL}\textbf{V}_{s,KL}^T,$\\ 
$\textbf{H}_{parent,L1}\textbf{H}_{s,L1}^T, ~~~~\textbf{H}_{parent,KL}\textbf{V}_{s,KL}^T$
\end{center} 
Then, the distance comparison is executed as
\begin{eqnarray}
Comp_{qL}&=&-2(\textbf{V}_{qL,L1}\textbf{V}_{s,L1}^T)+\textbf{V}_{qL,KL}\textbf{V}_{s,KL}^T \nonumber\\
Comp_h&=&-2(\textbf{H}_{parent,L1}\textbf{H}_{s,L1}^T)+\textbf{H}_{parent,KL}\textbf{V}_{s,KL}^T\nonumber
\end{eqnarray}
\begin{eqnarray}
&&Comp_{qL}-Comp_h \\
&&=r_s(DL1^{L1}_{qL,s}-DL1^{L1}_{parent,s})+2(\epsilon_{parent}^{'}-\epsilon_{qL})\nonumber \\
&&+r_s(DKL^{LAB}_{qL,s}-DKL^{LAB}_{parent,s})+(\epsilon_{parent}^{'}-\epsilon_{qL})\nonumber \\
&&=r_s(Dis(\textbf{V}_{qL}, \textbf{V}_{s})-Dis(\textbf{H}_{parent},\textbf{V}_{s}))\nonumber \\
&&+3(\epsilon_{parent}^{'}-\epsilon_{qL})\nonumber
\end{eqnarray}
where $\textbf{V}_{qL}$ is the least closest vector to $\textbf{V}_{req}$ in priority queue $Queue$. As $r_s$ is a positive value and $r_s>>(\epsilon_{parent}^{'}-\epsilon_{qL})$, the sign of $Comp_{qL}-Comp_h$ is consistent with $Dis(\textbf{V}_{qL}, \textbf{V}_{s})-Dis(\textbf{H}_{parent},\textbf{V}_{s})$.

$\bullet$ In the $Queue$ push process (line 30-37), privacy-preserving distance comparison is needed to determine whether a new node shall be added. Specifically, given $\textbf{C}(\textbf{V}_{s,L1})$, $\textbf{C}(\textbf{V}_{Node,L1})$, $\textbf{C}(\textbf{V}_{qL},L1)$, $\textbf{C}(\textbf{V}_{s,KL})$, $\textbf{C}(\textbf{V}_{Node,KL})$, $\textbf{C}(\textbf{V}_{qL},KL)$, the cloud use type-1 distance comparison in $PL1C-RF$ and $PKLC-RF$ to perform distance comparison as
\begin{eqnarray}
Comp_{Node}&=&-2(\textbf{V}_{Node,L1}\textbf{V}_{s,L1}^T)+\textbf{V}_{Node,KL}\textbf{V}_{s,KL}^T \nonumber\\
Comp_{qL}&=&-2(\textbf{V}_{qL,L1}\textbf{V}_{qL,L1}^T)+\textbf{V}_{cur,KL}\textbf{V}_{s,KL}^T\nonumber
\end{eqnarray}
\begin{eqnarray}
&&Comp_{Node}-Comp_{qL} \\
&&=r_s(DL1^{L1}_{Node,s}-DL1^{L1}_{qL,s})+2(\epsilon_{qL}-\epsilon_{Node})\nonumber \\
&&+r_s(DKL^{LAB}_{Node,s}-DKL^{LAB}_{qL,s})+(\epsilon_{qL}-\epsilon_{Node})\nonumber \\
&&=r_s(Dis(\textbf{V}_{Node}, \textbf{V}_{s})-Dis(\textbf{V}_{qL},\textbf{V}_{s}))\nonumber \\
&&+3(\epsilon_{qL}-\epsilon_{Node})\nonumber
\end{eqnarray}

To this end, the cloud is able to perform all operations required by a RKDF search in the privacy-preserving manner, and obtain a $Queue$ of nodes that stores data of top related images to the request. The cloud returns distance comparison candidates (type-1 distance) $Comp_i, i\in Queue$ as well as corresponding encrypted keywords back to the user.

\subsubsection{Final Keyword Selection}\label{s:finalkeywordprocessing} 
The user first decrypts encrypted keywords and obtains $K_{i,t}, i\in Queue$, where $K_{i,t}$ is the $t$-th pre-annotated keyword in image $I_i$. Then, the user computes distances $Dis(\textbf{V}_i,\textbf{V}_s), i\in Queue$ as 
\begin{eqnarray}\label{e:actdis}
&Dis(\textbf{V}_i,\textbf{V}_s)=(2r+\sum_{j=1}^{m_{L1}} v^2_{s,L1,j}) + \frac{Comp_i}{r_s}  \\
&=(2r+\sum_{j=1}^{m_{L1}} v^2_{s,L1,j})+\frac{-2(\textbf{V}_{i,L1}\textbf{V}_{s,L1}^T)+\textbf{V}_{i,KL}\textbf{V}_{s,KL}^T}{r_s} \nonumber
\end{eqnarray}
 To achieve higher accuracy in keywords selection, we consider that keywords in images that have smaller distance to the requested one are more relevant. Thus, we define a real-time weight $W_{t}$ for each keyword based on distances $Dis(\textbf{V}_i,\textbf{V}_s)$ as
\begin{flalign}\label{e:finalweight}
&W_{I_i}= 1-\frac{Dis(\textbf{V}_i,\textbf{V}_s)}{\sum_{i\in RST} Dis(\textbf{V}_i,\textbf{V}_s)}  \\
&W_{t} = \sum W_{I_i}, ~for~I_i~contains~K_{i,t} 
\end{flalign}
Specifically, we first figure out the weight $W_{I_i}$ of each image according to their distance-based similarity. As our definition in Eq.\ref{e:finalweight}, images with smaller distance will receive a larger weight value. Then, considering the same keyword can appear in multiple images, the final weight $W_{t}$ of a keyword $K_{i,t}$ is generated by adding weights of images that contain this keyword. Finally, the user selects keywords for his/her image according to their ranking of weight $W_{t}$.

\begin{figure*}[!ht]
\begin{center}
\includegraphics[height=3.85cm]{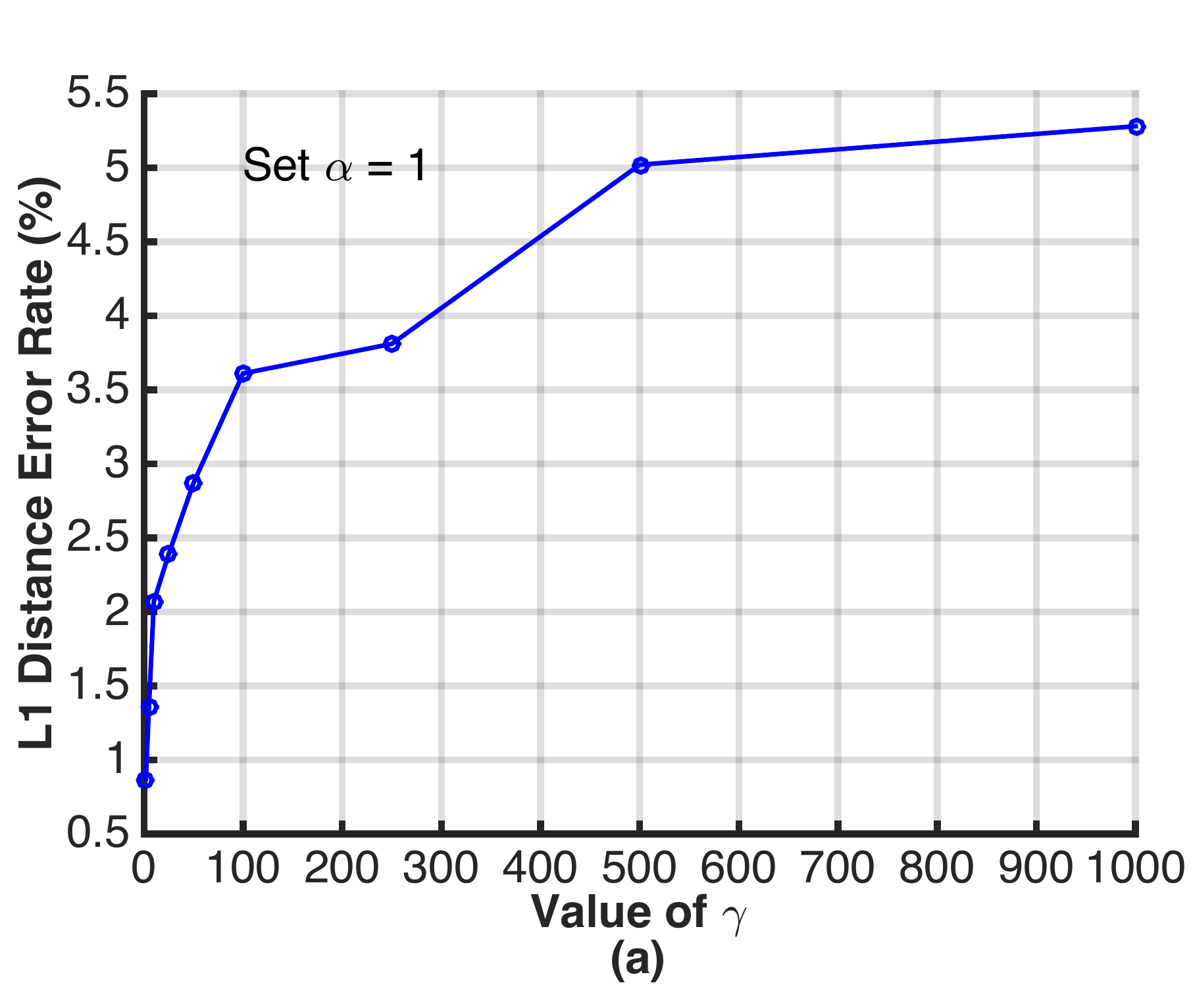}\includegraphics[height=3.85cm]{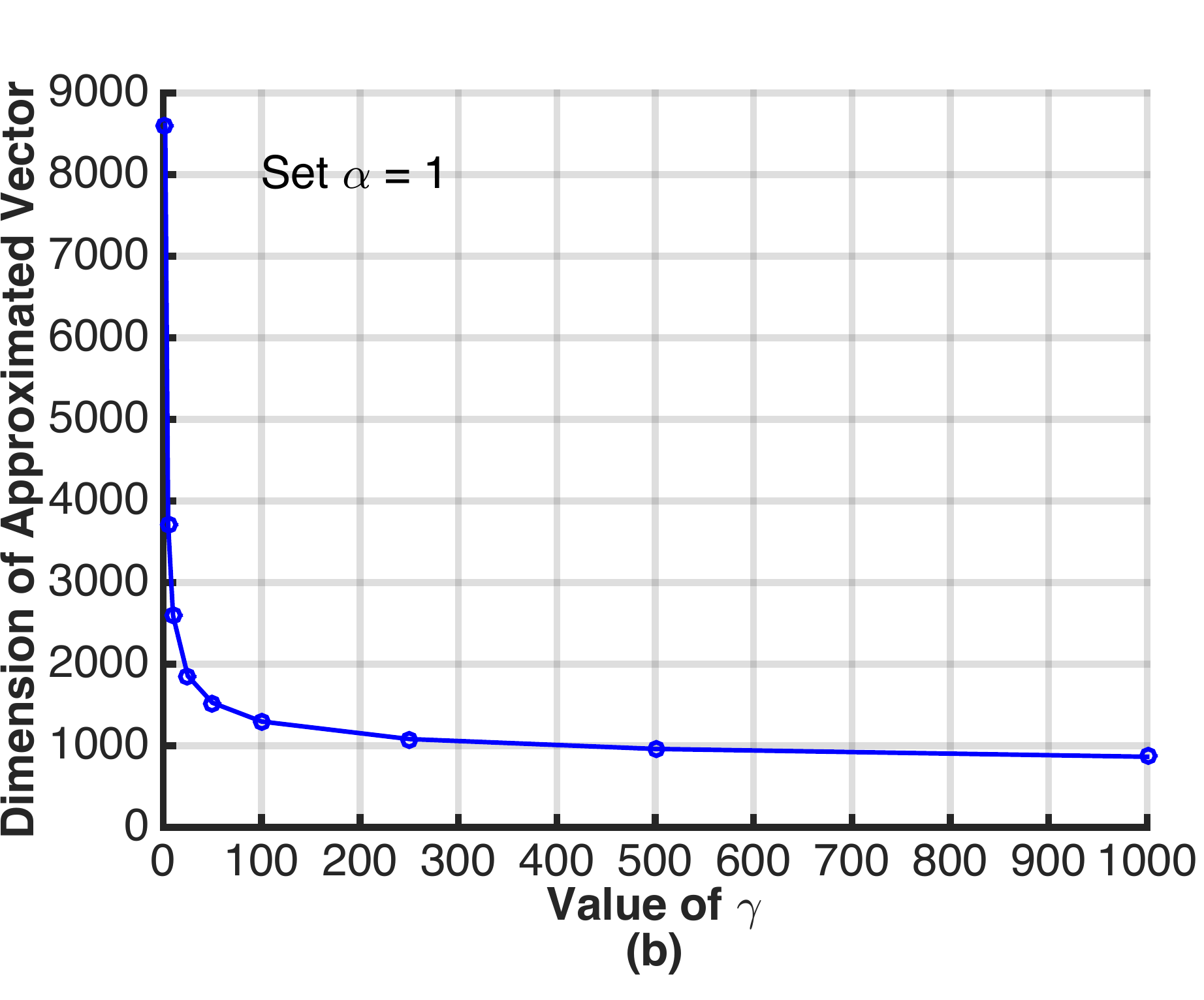}\includegraphics[height=3.85cm]{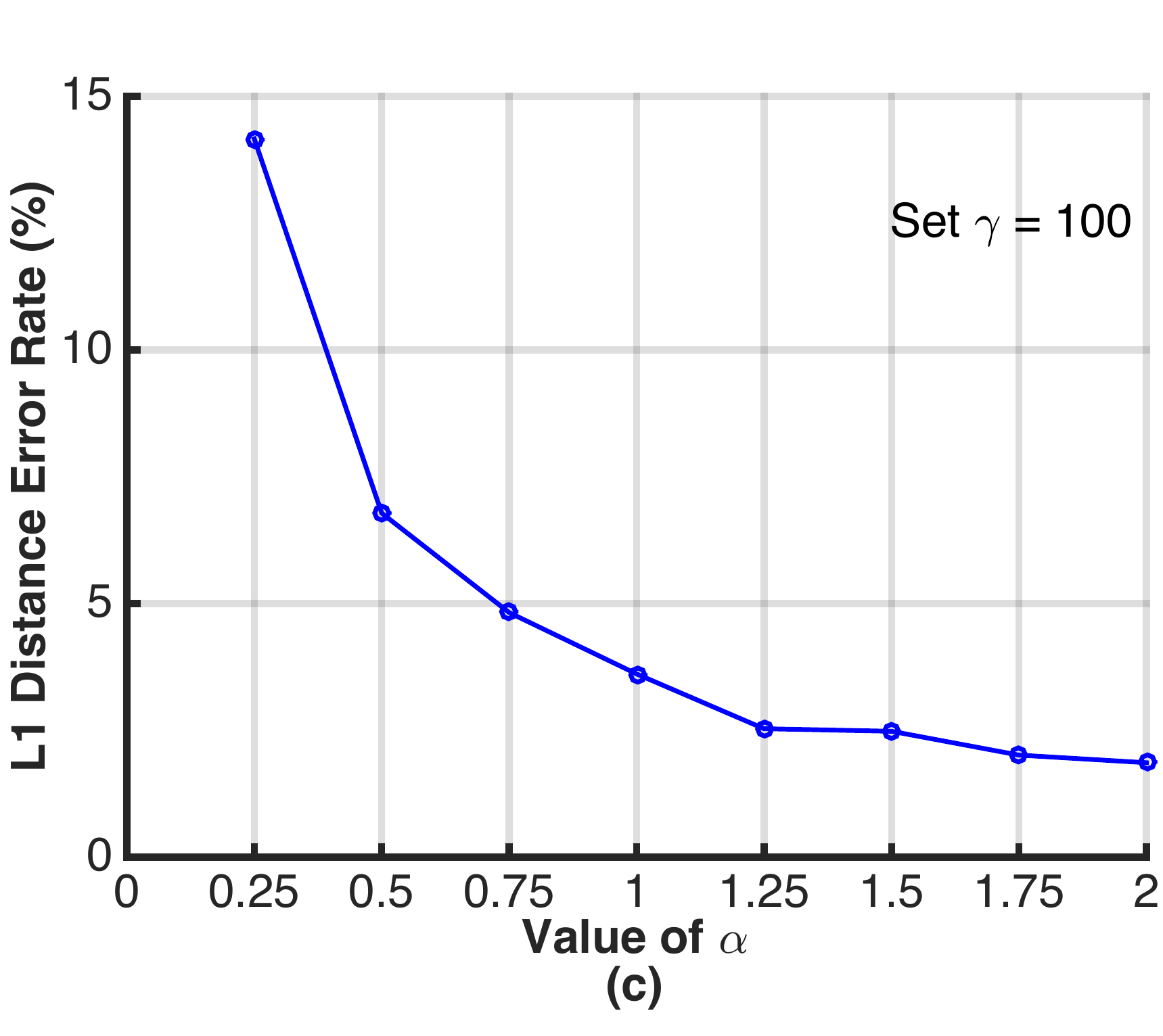}\includegraphics[height=3.85cm]{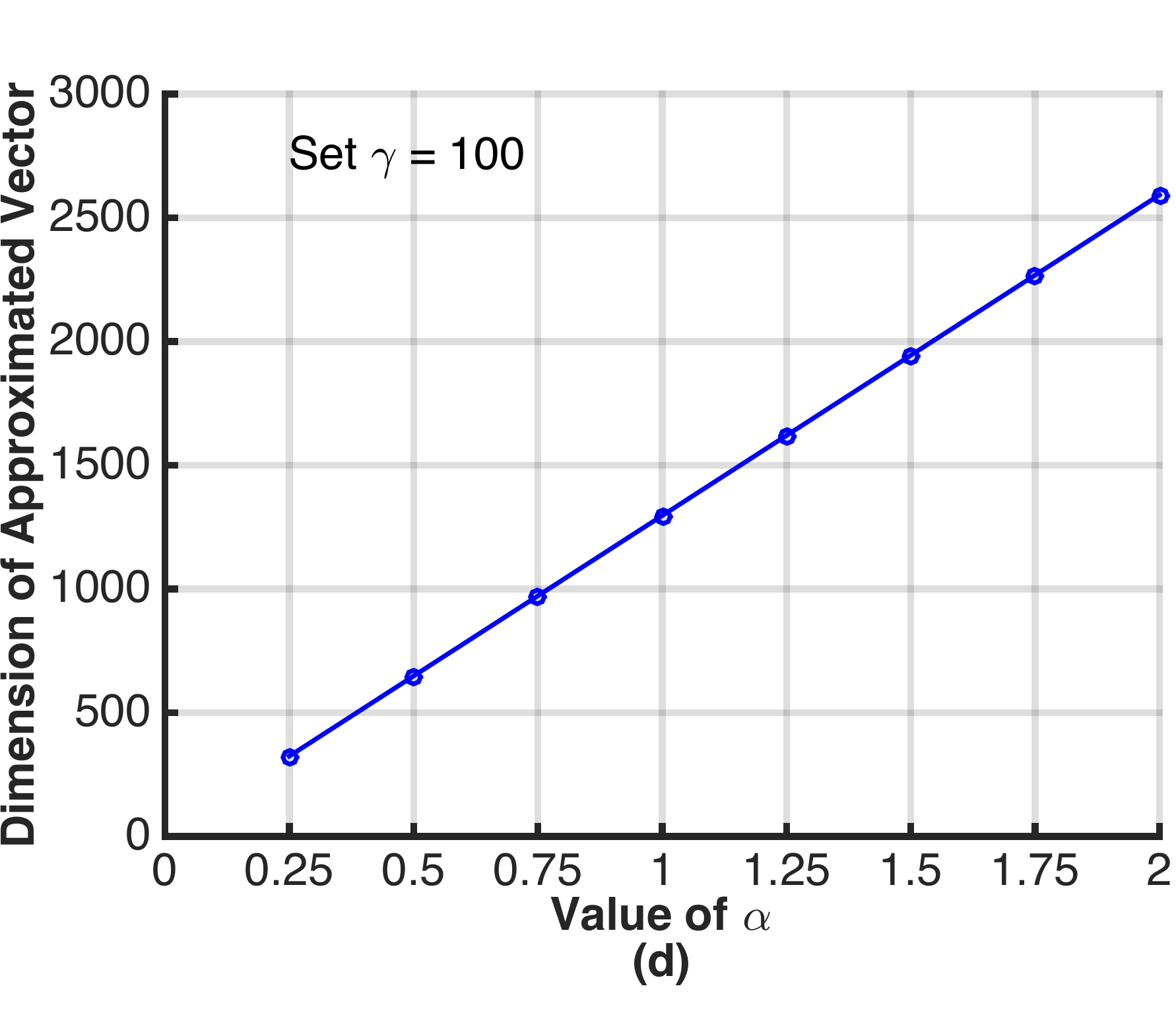}
\end{center}
\vspace{-5mm}
\centering
\caption{Error rate of Approximation and Dimension of Approximated Vector ($PCA-32$)}\label{f:alphagamma}
\vspace{-5mm}
\end{figure*}

\begin{algorithm}\label{a:kd-forest-search}
\small
\caption{Privacy-preserving RKDF Search}
  \SetKwInOut{Input}{Input}
  \SetKwInOut{Output}{Output}
  \Input{Encrypted Search Request (Req) for $\textbf{V}_s$, Encrypted RKDF with a set of Trees $\{T_k\}$, approximation power $\mathcal{AP-X}$}
  \Output{Encrypted Nodes Associated with Top Related Images to the Request.}
  Initialization $Queue$ = [], $Path$ = [] (Searched Path), $Vis$ = [] (Visited Nodes), $Node_k$=$T_k$.root; \\
  Each tree $T_k$ executes topDownTraversal() and backTraceSearch() in parallel, $Queue$ and $Vis$ are shared among all trees; \\
  \SetKwFunction{FTD}{topDownTraversal}
  \SetKwProg{Fn}{Function}{:}{}
  \Fn{\FTD{Req, $Node_k$}}{
  		\If{$Node_k$ is not null}{
  			return;
  		}
        $\textbf{V}_{i} \xleftarrow{} Node_k.\textbf{V}_{i}$; \\
        \eIf{$OPE(\textbf{V}_{s}[s_i]) \leq OPE(\textbf{V}_{i}[\textit{s}_{i}])$}{
          $Node_k$ = topDownTraversal($Node_k$.left-child); \\
        }{
          $Node_k$ = topDownTraversal($Node_k$.right-child); \\
        }

        \If{$Node_k \not \in Vis$}{
          $Vis.push(Node_k)$; \\
          $Queue.push(Node_k)$; \\
          }

        $Path.push(Node_k)$; \\
        return $Node_k$; \\
  }

  \SetKwFunction{FBT}{backTraceSearch}
  \SetKwProg{Fn}{Function}{:}{}
  \Fn{\FBT{Req, $Node_k$}}{
        \If{$Vis.length() > \mathcal{AP-X} \times$ Nodes Number}{
          return $Queue$; \\
        }
        \If{$Path$ is not null}{
          $\textit{parent} \xleftarrow{} Path.pop()$; \\
        }
        \If{$parent \not \in Vis$}{
            $Vis.push(\textit{parent})$; \\
            $Queue.push(parent)$; \\
        }
        {//Privacy-preserving distance comparison is achieved by $PL1C-RF$ and $PKLC-RF$, $\textbf{V}_{qL}$ is the least closest vector to $\textbf{V}_{s}$} in $Queue$\\
        \eIf{$Dis(\textbf{V}_{qL}, \textbf{V}_{s}) < Dis(\textbf{H}_{parent},\textbf{V}_{s}))$}{
          backTraceSearch(Req, $parent$); \\
        }{
          $Node_k$ = topDownTraversal$(Req, Node_k.sibling)$; \\
        }
  return $Queue$; \\
  }
  \SetKwFunction{FBT}{Queue.push}
  \SetKwProg{Fn}{Function}{:}{}
  \Fn{\FBT{$Node$}}{
  	//Each $Node_q$ in $Queue$ are ordered by $Dis(\textbf{V}_{s}, \textbf{V}_{Node_q})$ \\
  	\eIf{$Queue.length() < $ Defined Size $L$}{
  	Add $Node$ into $Queue$ by order;
  	}{
  		\If{$Node_{qL}$ in $Queue$ has $Dis(\textbf{V}_{s}, \textbf{V}_{Node})<Dis(\textbf{V}_{s}, \textbf{V}_{qL})$}{
  		Remove $Node_{qL}$ from $Queue$;\\
  		Add $Node$ into $Queue$ by order;
  		}
  	}
  }
\end{algorithm}

\section{Security Analysis}\label{s:securityanalysis}
In CPAR, we have the following privacy related data: feature vectors $\{\textbf{V}_{i,L1},\textbf{V}_{i,KL}\}_{1\leq i\leq n}$, hyperplane projected vectors $\textbf{H}_{i,L1}$, $\textbf{H}_{i,KL}$ of each non-leaf node associated with $\textbf{V}_{i,L1}$, $\textbf{V}_{i,KL}$, the \textit{split} field element of each non-leaf node, keywords of image $I_i$ in the pre-annotated dataset, and feature vectors $\textbf{V}_{s,L1}$, $\textbf{H}_{s,L1}$, $\textbf{V}_{s,KL}$ of the image requested for annotation. As keywords are encrypted using standard AES encryption, we consider them secure against the cloud server as well as outside adversaries. For the \textit{split} field element of each non-leaf node, it is encrypted using the order-preserving encryption \cite{OPE,OPE3}, which has been proved to be secure. With regards to $\textbf{V}_{i,L1}$, $\textbf{H}_{i,L1}$, $\textbf{V}_{i,KL}$, $\textbf{H}_{i,KL}$, $\textbf{V}_{s,L1}$, $\textbf{H}_{s,L1}$ $\textbf{V}_{s,KL}$, they are encrypted using the encryption scheme of IVE \cite{zhou2014efficient} after pre-processing as presented in our $PL1C-RF$ and $PKLC-RF$ schemes. The IVE scheme \cite{zhou2014efficient} has been proved to be secure based on the well-known Learning with Errors (LWE) hard problem \cite{LWE}. Thus, given the ciphertexts $\textbf{C}(\textbf{V}_{i,L1})$, $\textbf{C}(\textbf{H}_{i,L1})$, $\textbf{C}(\textbf{V}_{i,KL})$, $\textbf{C}(\textbf{H}_{i,KL})$, $\textbf{C}(\textbf{V}_{s,L1})$, $\textbf{C}(\textbf{H}_{s,L1})$, $\textbf{C}(\textbf{V}_{s,KL})$ only, it is computational infeasible for the cloud server or outside adversaries to recover the corresponding feature vectors.

\subsection{Security of Outsourcing $\textbf{S}_{L1}^T\textbf{S}_{s,L1}$, $\textbf{S}_{L1}^{'T}\textbf{S}_{s,L1}^{'}$ and $\textbf{S}_{KL}^T\textbf{S}_{s,KL}$} 

As $\textbf{S}_{L1}^T\textbf{S}_{s,L1}$, $\textbf{S}_{L1}^{'T}\textbf{S}_{s,L1}^{'}$, and $\textbf{S}_{KL}^T\textbf{S}_{s,KL}$ are used in the same manner, we use $\textbf{S}^T\textbf{S}_{s}$ to denote them for expression simplicity. Different from the original encryption algorithm of IVE, the user in CPAR also outsources $\textbf{S}^T\textbf{S}_s$ to the cloud besides ciphertexts. As all elements in $\textbf{S}$ and $\textbf{S}_s$ are randomly selected, elements in their multiplication $\textbf{S}^T\textbf{S}_s$ have the same distribution as these elements in $\textbf{S}$ and $\textbf{S}_s$ \cite{Crpto-book-cp11}. Thus, given $\textbf{S}^T\textbf{S}_s$, the cloud server is not able to extract $\textbf{S}$ or $\textbf{S}_s$ directly and use them to decrypt ciphertexts. By combining $\textbf{S}^T\textbf{S}_s$ with ciphertexts $\textbf{C}(\textbf{V}_{i,L1})$ and $\textbf{C}(\textbf{V}_{s,L1})$ (same as that for $\textbf{C}(\textbf{H}_{i,L1})$, $\textbf{C}(\textbf{H}_{s,L1})$, $\textbf{C}(\textbf{V}_{i,KL})$, $\textbf{C}(\textbf{H}_{i,KL})$ and $\textbf{C}(\textbf{V}_{s,KL})$), the cloud can obtain
\begin{eqnarray}\label{e:security}
\textbf{S}^T\textbf{S}_s\textbf{C}(\textbf{V}_{i,L1})&=&\textbf{S}^T\textbf{S}_s\textbf{S}^{-1}(w\textbf{V}_{i,L1}+\textbf{e}_i)^T\nonumber\\
\textbf{S}^T\textbf{S}_s\textbf{C}(\textbf{V}_{s,L1})&=&\textbf{S}^T\textbf{S}_s\textbf{S}_s^{-1}(w\textbf{V}_{s,L1}+\textbf{e}_s)^T \nonumber \\
&=&\textbf{S}^T(w\textbf{V}_{s,L1}+\textbf{e}_s)^T  \nonumber
\end{eqnarray} 
From the above two equations, it is clear that the combination of $\textbf{S}^T\textbf{S}_s$, $\textbf{C}(\textbf{V}_{i,L1})$ and $\textbf{S}^T\textbf{S}_s$, $\textbf{C}(\textbf{V}_{s,L1})$ only transfer them to the ciphertexts of $\textbf{V}_{i,L1}$ and $\textbf{V}_{s,L1}$ that encrypted using the IVE scheme with new keys $\textbf{S}^T\textbf{S}_s\textbf{S}^{-1}$ and $\textbf{S}^T$ respectively. As $\textbf{S}^T\textbf{S}_s\textbf{S}^{-1}$ and $\textbf{S}^T$ are random keys and unknown to the cloud, recovering $\textbf{V}_{i,L1}$, $\textbf{V}_{s,L1}$ from $\textbf{S}^T\textbf{S}_s\textbf{C}(\textbf{V}_{i,L1})$, $\textbf{S}^T\textbf{S}_s\textbf{C}(\textbf{V}_{s,L1})$ still become the $LWE$ problem as proved in ref \cite{zhou2014efficient}. To this end, $\textbf{S}^T\textbf{S}_s$ only helps the cloud perform distance comparison in CPAR, but does not bring additional advantages to recover feature vectors compared with the given ciphertexts only scenario.

\subsection{Request Unlinkability}
The request unlinkability in CPAR is guaranteed by the randomization for each request. Specifically, each query request $\{\textbf{V}_{s,L1},\textbf{H}_{s,L1},\textbf{V}_{s,KL}\}$ is element-wise obfuscated with different random error terms $\textbf{e}_s$, $\textbf{e}_{s}^{'}$ and random number $r_s$ during the encryption, which makes the obfuscated $\textbf{V}_{s,L1},\textbf{H}_{s,L1},\textbf{V}_{s,KL}$ have the same distribution as in these random values $\textbf{e}_s$, $\textbf{e}_{s}^{'}$ and $r_c$ \cite{Crpto-book-cp11}. Thus, by changing $\textbf{e}_s$, $\textbf{e}_{s}^{'}$ and $r_c$ during the encryption of different requests, CPAR outputs different random ciphertexts, even for requests generated from the same image.

\section{Evaluation}\label{s:evaluation}
To evaluate the performance of CPAR, we implemented a prototype using Python 2.7. In our implementation, Numpy \cite{developers2013numpy} is used to support efficient multi-dimension array operations. OpenCV \cite{bradski2000opencv} is used to extract the color-space features of the images and build the filter kernels to generate the Gabor filter results. Pywt \cite{pywt} is adopted to perform Haar wavelet and get the corresponding Haar results. Sklearn \cite{scikit-learn} is used to perform the PCA transformation. FLANN library \cite{muja2014scalable} is used to act as the non-privacy randomized kd-forest for comparison. We use the well-known IAPR TC-12 \cite{IAPR} as the pre-annotated dataset, which contains 20,000 annotated images and the average number of keywords for each image is 5.7. All tests are performed on a 3.1 GHz Intel Core i7 Macbook Pro with OS X 10.14.2 installed as \textit{User} and a Microsoft Azure cloud E4-v3 VM with Ubuntu 18.04 LTS installed as \textit{Cloud Server}.

In the rest of this section, $n$ is the total number of images in the pre-annotated dataset, $m_{L1}$ is the dimension of pre-processed feature vectors $\textbf{V}_{i,L1}$, $m_{KL}$ is the dimension of pre-processed feature vectors $\textbf{V}_{i,KL}$ and their corresponding hyperplane projected vectors $\textbf{H}_{i,KL}$, $m_{L1}^{'}$ is the dimensions of hyperplane projected vector $\textbf{H}_{i,L1}$. We also use $DOT_{m}$ to denote a dot product operation between to two $m$-dimensional vectors. $\mathcal{AP}-X$ is used to denote the approximation power during the RKDF search, which indicates $X\%$ of the nodes will be checked in each tree of RKDF. $PCA-X$ is used to denote the strength of $PCA$ transformation applied to $\textbf{V}_{i,{H}}$ and $\textbf{V}_{i,{HQ}}$ in $\textbf{V}_{i,L1}$, which compresses their dimensions from 4096 to $\frac{4096}{X}$. $PCA-128$, $PCA-64$, $PCA-32$, $PCA-16$, and $PCA-8$ are evaluated in our experiments to balance the efficiency and accuracy of CAPIA.



In our evaluation, we first provide numerical analysis as well as experimental evaluation for each stage of CPAR. Then, we compare CPAR with CAPIA proposed in ref \cite{tian2017capia} in terms of efficiency and accuracy.

\subsection{System Parameter Selection}\label{ss:setupandenc}
To perform the one-time setup in CPAR, the user pre-processes feature vectors of each image in the pre-annotated image dataset. Specifically, the user first performs JL-Lemma based approximation over $\textbf{V}_{i,L1}$ to make them compatible with our $PL1C-RF$. As discussed in Section \ref{ss:PL1C-RF}, there is a trade-off between the approximation accuracy of $L_1$ distance and length of the approximated vector that determines efficiency of follow up privacy-preserving operations. To balance such a trade-off, we evaluate different parameters for approximation as shown in Fig.\ref{f:alphagamma} (a)-(d). According to our results, we suggest to set $\alpha=1$ and $\gamma=100$ which introduces $3.61\%$ error rate for $L_1$ distance computation, and extends the dimension of $\textbf{V}_{i,L1}$ from $864$ to $1296$ under the setting of $PCA-32$. Specifically, the error rate drops fast when $\alpha<1$ and becomes relative stable when $\alpha>1$. Meanwhile, the dimension of the approximated vector increases linearly to the value of $\alpha$. With regards to $\gamma$, the dimension of the approximated vector becomes relative stable when $\gamma>100$, however, the error rate still increases when $\gamma>100$. 

\begin{figure}[ht]
\begin{center}
\includegraphics[height=5cm]{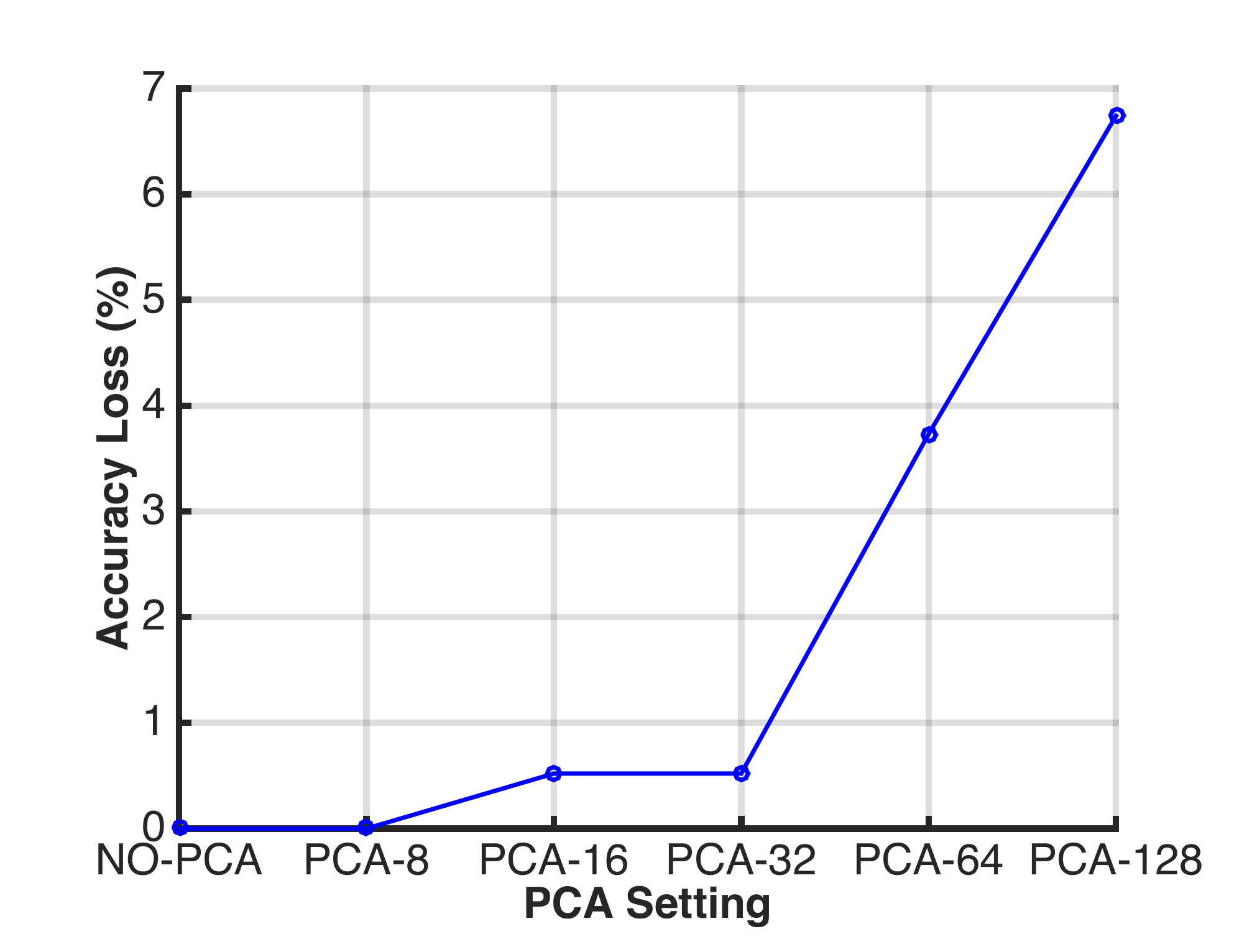}
\end{center}
\caption{Accuracy Loss with Different PCA Settings}\label{f:pca}
\end{figure} 

With regards to the selection of $PCA$ parameter, it is clear that better efficiency of CPAR will be achieved by increasing the strength of $PCA$. However, the stronger $PCA$ setting will also cause accuracy loss due to the loss of information during the compression. To balance the efficiency and accuracy, we evaluate of accuracy loss of annotation with different PCA setting. Compared with the $No-PCA$ setting, Fig.\ref{f:pca} shows the accuracy loss for $PCA-8$, $PCA-16$, and $PCA-32$ are stable and bounded in $0.5\%$. Differently, $PCA-64$ and $PCA-128$ rapidly raise the accuracy loss. Therefore, $PCA-32$ is adopted by CPAR.

\subsection{RKDF Construction and Encryption}\label{ss:rkdf-enc}
To construct an encrypted RKDF, the user first constructs an unencrypted RKDF using 20,000 pre-annotated images, and then replaces data of each node in the RKDF with their corresponding ciphertexts. The construction of an unencrypted RKDF with 10 kd-trees costs 28.56 seconds. Then, for the pre-processed feature vectors $\textbf{V}_{i,L1}$ and $\textbf{V}_{i,KL}$ of each image, the user can encrypt them using $PL1C-RF$ and $PKLC-RF$ with $(m_{L1})DOT_{m_{L1}}$ and $(m_{KL})DOT_{m_{KL}}$ operations respectively, which costs 8.4ms in total in our implementation. If an image is associated with a non-leaf node in any tree of the RKDF, encryption for the hyperplane projected vectors $\textbf{H}_{i,L1}$ and $\textbf{H}_{i,KL}$ with $(m_{L1}^{'})DOT_{m_{L1}^{'}}$ and $(m_{KL})DOT_{m_{KL}}$ operations respectively, which costs 54.7ms in total. In addition, for each non-leaf node, an order-preserving encryption is needed for the \textit{split} field, each of which costs 1.4ms. Therefore, to build a 10-tree encrypted RKDF with a 20,000 pre-annotated image dataset, it takes 74.78 minutes in our implementation. It is noteworthy that the encrypted RKDF construction is one-time offline cost, which does not impact the performance of later on real-time privacy-preserving image annotation.  



\subsection{Real-time Image Annotation}\label{ss:real-time-anno}
\textit{Request Generation}: To annotate a new image in a privacy-preserving manner, the user pre-processes and encrypts its feature vectors $\textbf{V}_{s,L1}$ and $\textbf{V}_{s,KL}$ using $PL1C-RF$ and $PKLC-RF$. Specifically, the encryption of $\textbf{V}_{s,L1}$ requires $(m_{L1})DOT_{m_{L1}}+(m_{L1}^{'})DOT_{m_{L1}^{'}}$ for shown in Fig.\ref{f:pl1c-Rf}, and the encryption of $\textbf{V}_{s,KL}$ requires $(m_{KL})DOT_{m_{KL}}$ operations as shown in Fig.\ref{f:pklc-Rf}. In addition, for each element $sf_j$ in the \textit{split} field element set $\mathcal{SF}$ with size of 348 in our implementation, order-preserving encryption are executed for $\textbf{V}_{s}[sf_j]$. As a result, the encrypted request can be efficiently generated with only 534.16ms. 

\begin{figure}[!ht]
\begin{center}
\includegraphics[height=5cm]{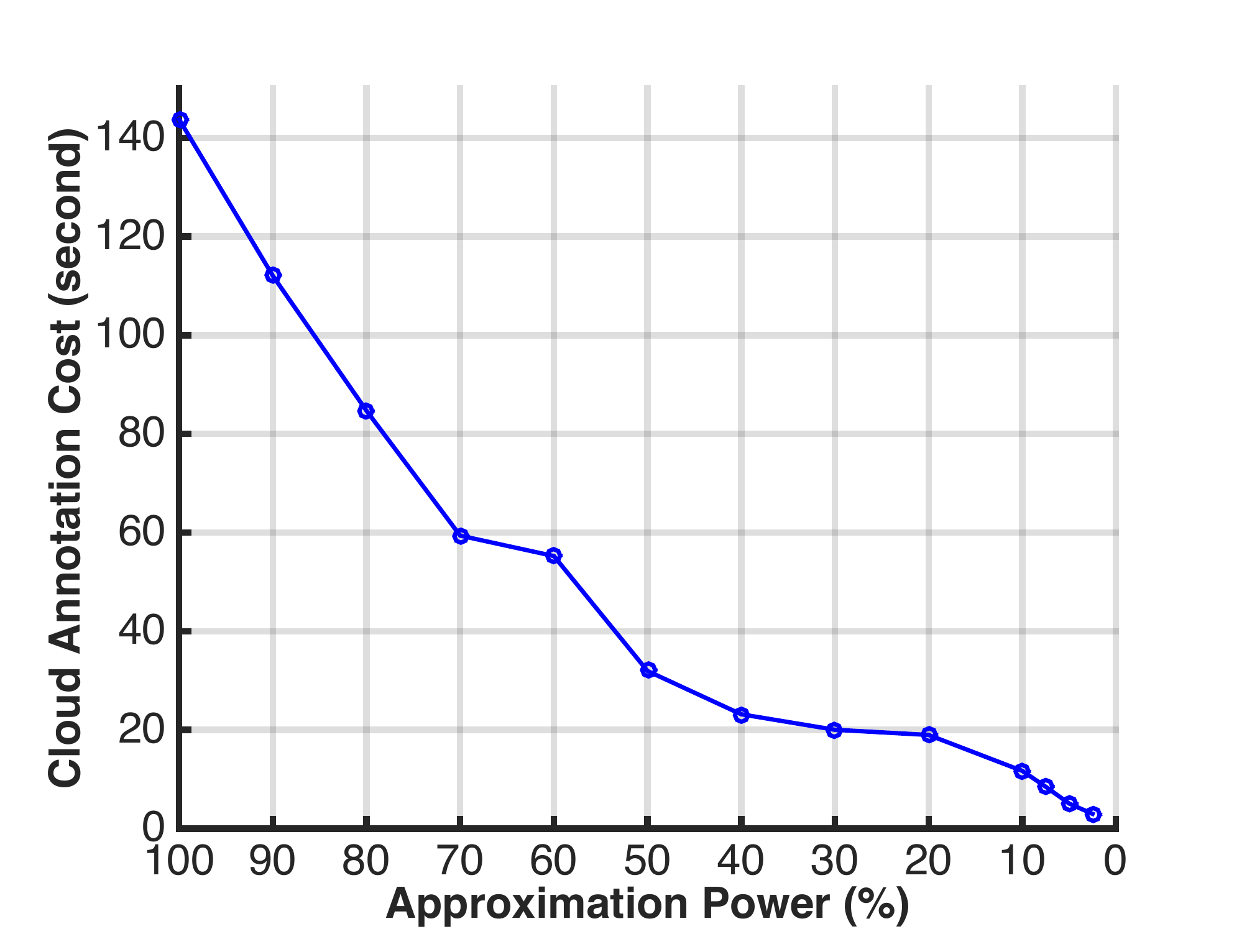}
\end{center}
\caption{Privacy-preserving Annotation Cost on Cloud with Different Approximation Power}\label{f:eff-ap}
\end{figure} 

\begin{figure}[!ht]
\begin{center}
\includegraphics[height=5cm]{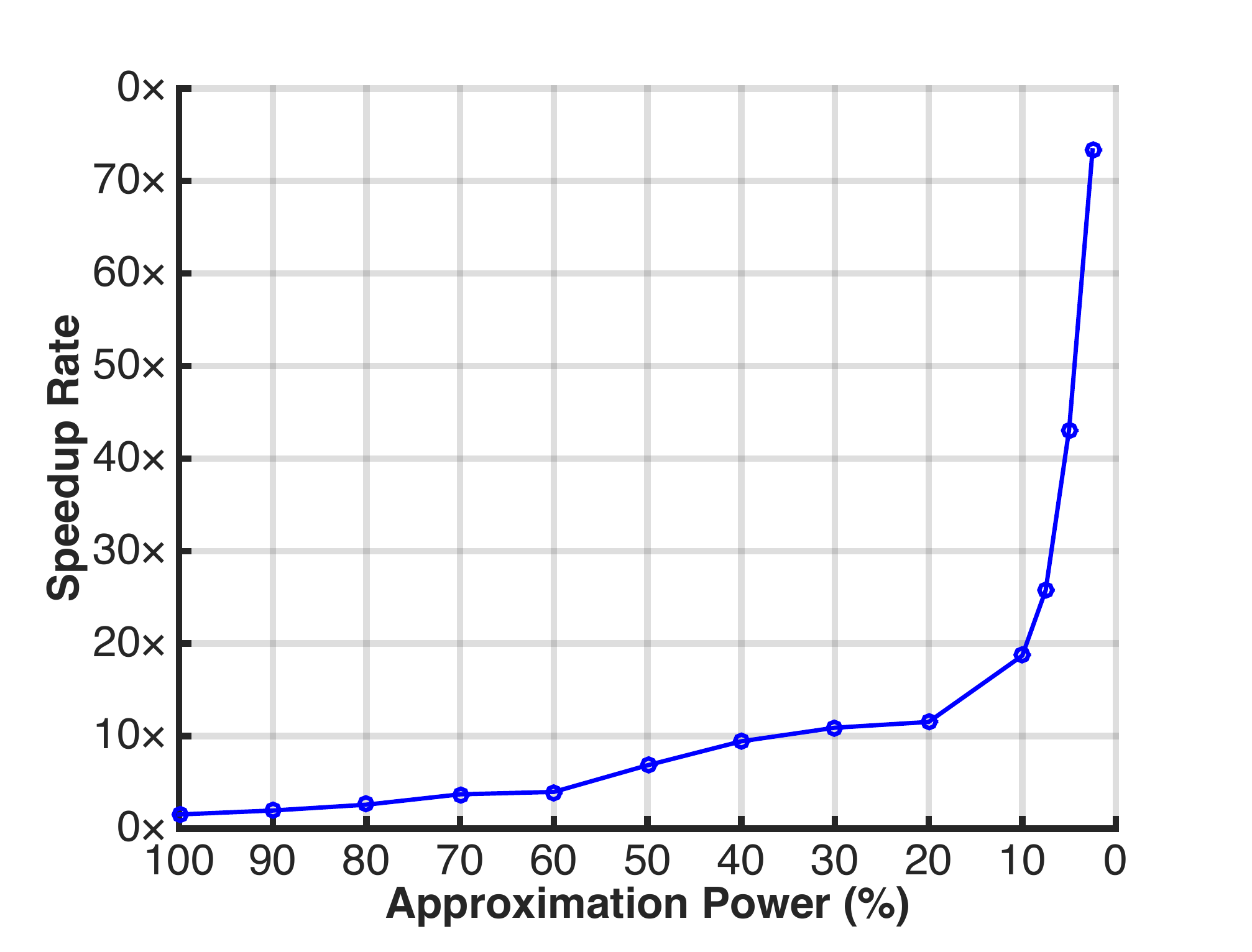}
\end{center}
\caption{Speedup Rate with Different Approximation Power}\label{f:speedup}
\end{figure}

\textit{Privacy-preserving Annotation on Cloud}: On receiving the encrypted request, the cloud performs privacy-preserving RKDF search with top-down traversal, back trace search, and queue push processes. The top-down traversal only requires a direct comparison between the ciphertexts under order-preserving encryption, whose cost is negligible compared with the other two processes. In the back trace search, privacy-preserving type-2 distance comparison needs to the executed using $PL1C-RF$ and $PKLC-RF$. In particular, two comparison candidates $Comp_{qL}$ and $Comp_h$ are computed with $(m_{L1}+1)DOT_{m_{L1}}+(m_{KL}+1)DOT_{m_{KL}}$ operations and $(m_{L1}^{'}+1)DOT_{m_{L1}^{'}}+(m_{KL}+1)DOT_{m_{KL}}$ operations respectively. With regards to the queue push process, privacy-preserving type-1 distance comparison are executed using $PL1C-RF$ and $PKLC-RF$, which requires $2(m_{L1}+1)DOT_{m_{L1}}+2(m_{KL}+1)DOT_{m_{KL}}$ operations in total. Another important parameter that affects the search efficiency is the selection of approximation power $\mathcal{AP-X}$. As depicted in Fig.\ref{f:eff-ap}, by increasing the approximation power from $\mathcal{AP}-100$ to $\mathcal{AP-}2.5$, the privacy-preserving annotation using encrypted RKDF reduces from 143.72 seconds to 2.98 seconds. Compared with CAPIA \cite{tian2017capia} that requires 218.46 seconds for one privacy-preserving annotation on cloud and does not support approximate dataset checking, CPAR can significantly speed it up as depicted in Fig.\ref{f:speedup}.  


\textit{Final Keyword Selection}: This process only involves AES decryption and the weights generation that only requires a small number of additions. As a result, the final keyword selection can be completed by the user within 318ms. 


\begin{figure}[!ht]
\begin{center}
\includegraphics[height=5cm]{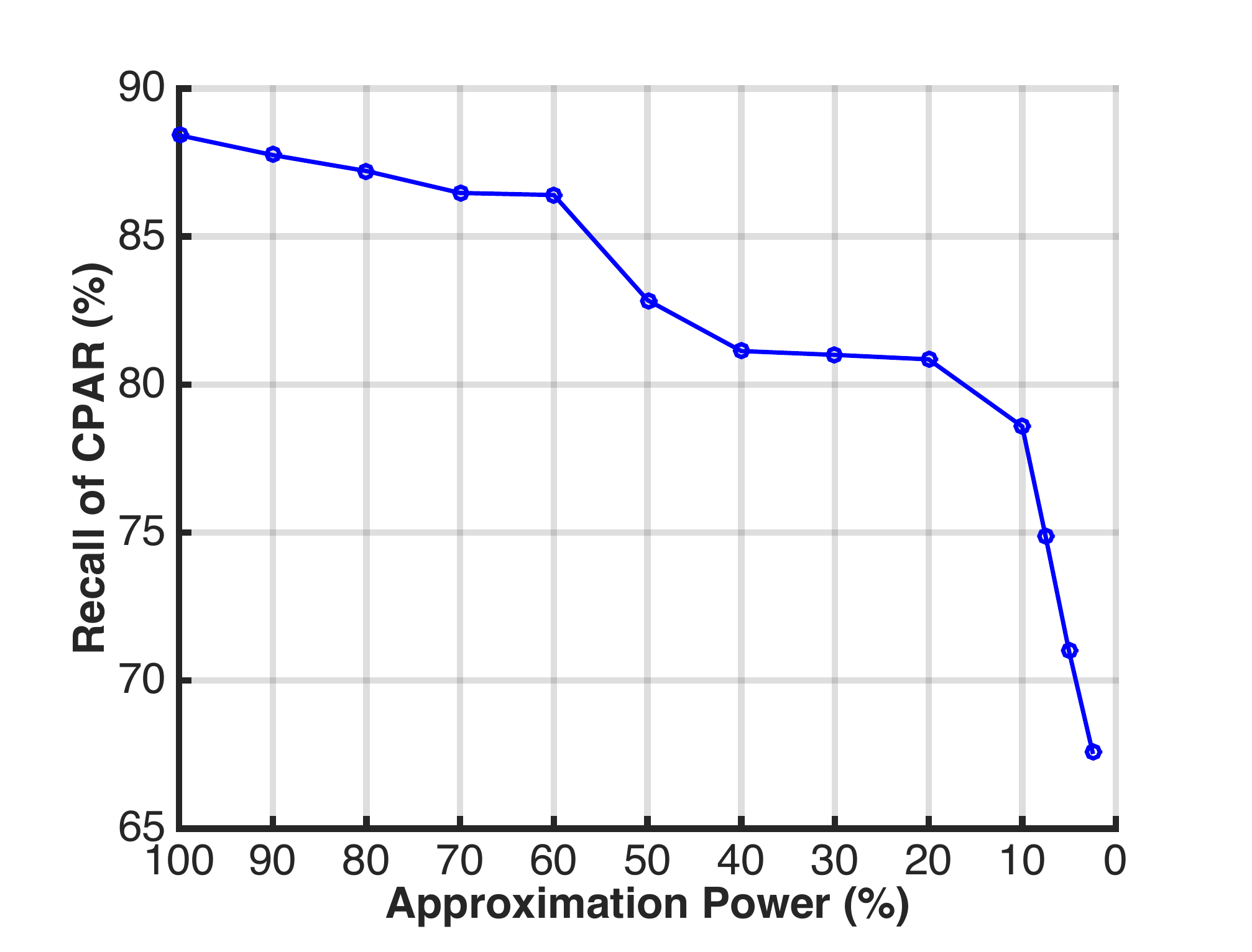}
\end{center}
\caption{Accuracy (Recall) of CPAR with Different Approximation Power}\label{f:app-recall}
\end{figure} 

\begin{figure}[!ht]
\begin{center}
\includegraphics[height=5cm]{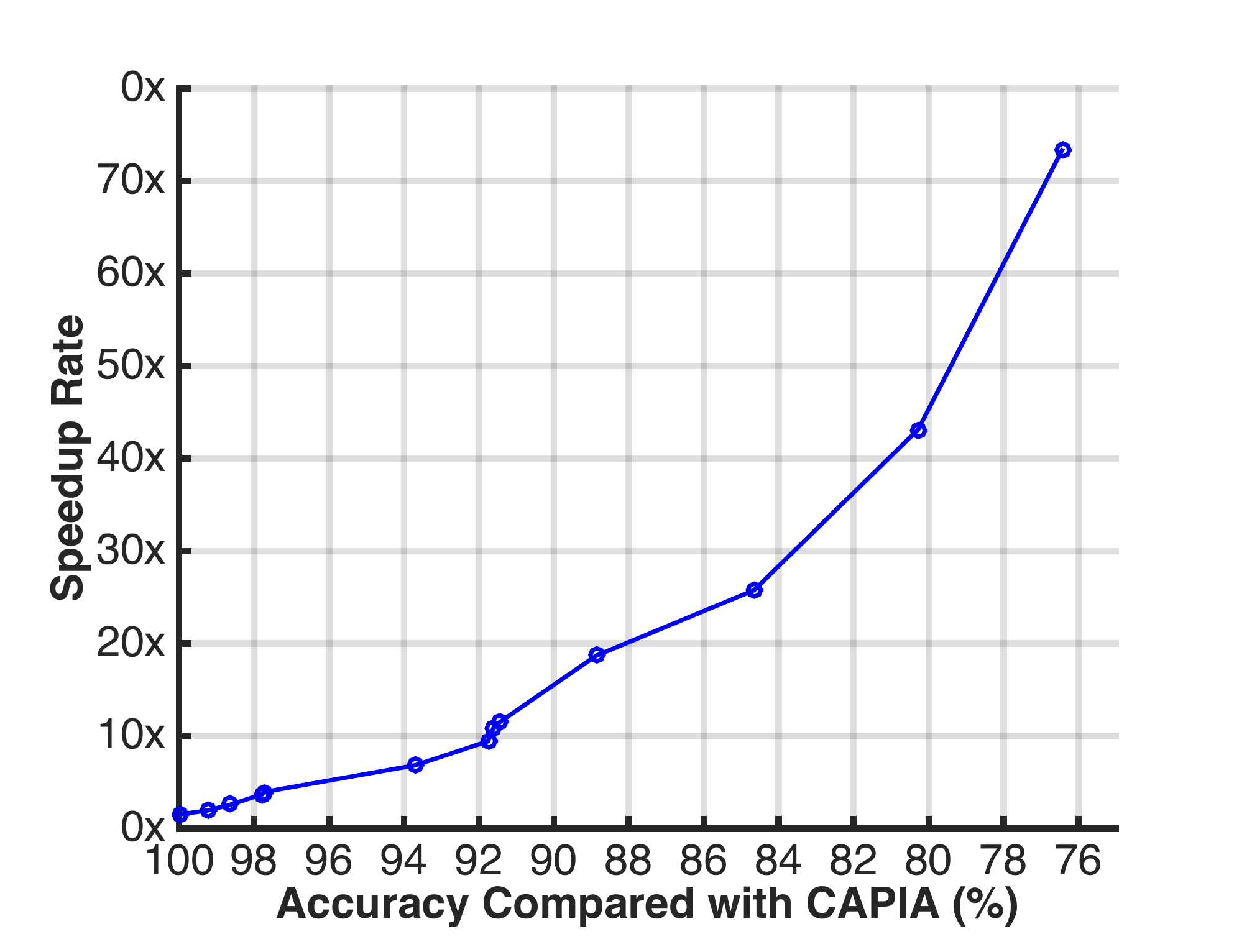}
\end{center}
\caption{Speedup rate of CPAR with Different Accuracy Compared with CAPIA }\label{f:speedup-recall}
\end{figure} 


\begin{table*}[ht]
  \caption{Sample Annotation Results} \label{t:sample}
  \center
  \begin{tabular}
      {|L{1.4cm}|c|c|c|c} \hline \centering Image & CPAR Annotation & CAPIA Annotation & Human Annotation \\
      \hline
      \parbox[c]{1em}{\includegraphics[height=11mm]{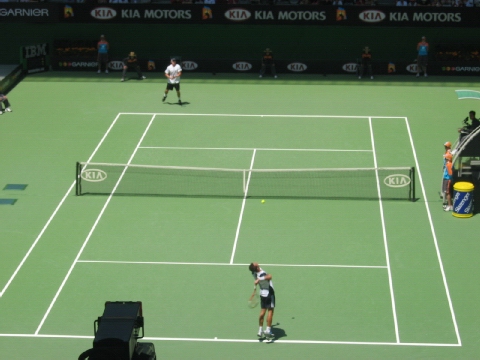}} & \begin{tabular}{@{}c@{}} \underline{\textbf{floor-tennis-court, man}}, grass \end{tabular} & \begin{tabular}{@{}c@{}} \underline{\textbf{floor-tennis-court, man}}, woman\end{tabular} & \begin{tabular}{@{}c@{}} floor-tennis-court, man \end{tabular} \\
      \hline
      \parbox[c]{1em}{\includegraphics[height=10mm]{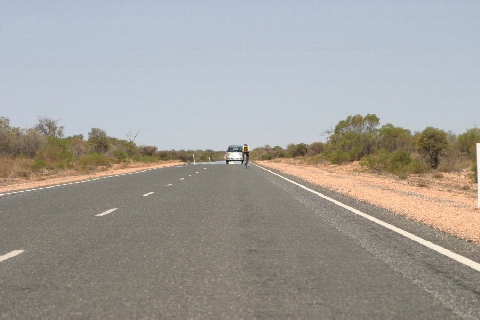}} & \begin{tabular}{@{}c@{}} \underline{\textbf{highway, sky-blue, trees, vegetation}}, \\ground, ship, sky, ocean, bush \end{tabular} & \begin{tabular}{@{}c@{}} \underline{\textbf{sky-blue, highway, vegetation}}, \\ground, bush, \underline{\textbf{trees}}, lake, ocean \end{tabular} & \begin{tabular}{@{}c@{}} highway, sky-blue, \\trees, vegetation \end{tabular} \\
      \hline
      \parbox[c]{1em}{\includegraphics[height=11mm]{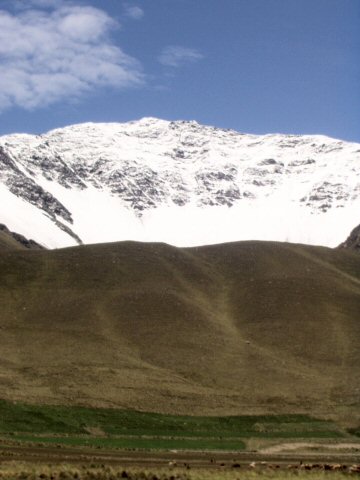}} & \begin{tabular}{@{}c@{}} group-of-persons, \underline{\textbf{ground, cloud}}, man, sky-light, \\\underline{\textbf{mountain}}, door, chair, floor-other, column \end{tabular} & \begin{tabular}{@{}c@{}} \underline{\textbf{cloud, sky-blue, ground}}, \\\underline{\textbf{mountain}}, horse man, road, \underline{\textbf{grass}} \end{tabular} & \begin{tabular}{@{}c@{}} ground, cloud, sky-blue, \\mountain, snow, grass \end{tabular} \\
      \hline
      \parbox[c]{1em}{\includegraphics[height=10mm]{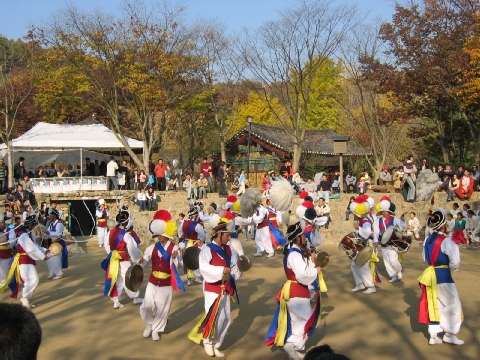}} & \begin{tabular}{@{}c@{}} \underline{\textbf{group-of-persons}}, hat, hill, cloud, \underline{\textbf{sky-blue}}, \\\underline{\textbf{ground}}, sky, fabric, couple-of-persons, grass \end{tabular} & \begin{tabular}{@{}c@{}} \underline{\textbf{group-of-persons, sky-blue, ground, trees}}, \\mountain, ruin-archeological, hat, cloud, hill \end{tabular} & \begin{tabular}{@{}c@{}} trees, ground, man, \\sky-blue, group-of-persons \end{tabular} \\
      \hline
  \end{tabular}
  \center
  In each cell of CPAR and CAPIA annotation results, ground-truth human annotation results are underlined and bold out. 
\end{table*}

\textbf{Accuracy}: To evaluate the accuracy of CPAR, we use the standard average \textit{recall} rates to measure the accuracy of keywords annotation. To be specific, by using $[K_1,K_2,\cdots,...K_y]$ to denote distinct keywords annotated with CPAR for a set of image annotation requests, the recall rate for each keyword $K_j$ and the average accuracy are defined as 
\begin{itemize}
  \item $recall_{K_j}=\frac{\#~of~images~assigned~K_j~correctly~by~CPAR}{\#~of~images~assigned~K_j~in~the~ground-truth}$
  \item $Accuracy=\frac{\sum_{j=1}^y recall_{K_j}}{y}$
\end{itemize}

In our evaluation, annotation requests for 50 different images are submitted, in which each requested image has two or more related images in the pre-annotated dataset. As shown in Fig.\ref{f:app-recall}, the accuracy of CPAR reduces from $88.42\%$ to $67.59\%$ when the approximation power increases from $\mathcal{AP}-100$ to $\mathcal{AP-}2.5$. Compared with CAPIA \cite{tian2017capia} our scheme achieves the same accuracy by setting the approximation power as $\mathcal{AP}-100$. 
While the increasing of approximation power reduces the accuracy of CPAR to some extent, it also boosts the efficiency significantly as shown in Fig.\ref{f:eff-ap}. Compared with CAPIA, Fig.\ref{f:speedup-recall} shows that CPAR can speed up CAPIA by 4$\times$, 11.5$\times$, 18.7$\times$, 25.8$\times$, 43.1$\times$ when achieving $97.7\%$, $91.4\%$, $88.9\%$, $84.7\%$, $80.3\%$ accuracy of CAPIA respectively. Therefore, CPAR can greatly promote the efficiency the of CAPIA while retaining comparable accuracy. To balance the efficiency speedup and annotation accuracy of CPAR, we suggest to set the approximation power as $\mathcal{AP}-10$, i.e. achieves $88.9\%$ accuracy of CAPIA with 18.7$\times$ speedup.

 In Table \ref{t:sample}, we present samples of automatically annotated images using CAPIA and CPAR with approximation power as $\mathcal{AP}-10$. In the last column we list the human annotation results (ground-truth) for comparison. On one hand, CPAR is highly possible to assign correct keywords to images compared with human annotation. This observation also confirms the high average recall rate of CPAR, since these ground-truth annotations are likely to be covered in CPAR. On the other hand, CPAR also introduces additional keywords that frequently appear together with these accurate keywords in top related images. These additional keywords are typically not directly included in human annotations, but are potentially related to correct keywords. Compared with CAPIA, CPAR only misses a small portion of ground-truth keywords due to the approximation strategy, which is consistence with our evaluation result in Fig.\ref{f:app-recall} and Fig.\ref{f:speedup-recall}. Overall, our evaluation results demonstrate that although CPAR cannot provide perfect keywords selection all the time compared with human annotation, it can still maintain comparable accuracy as CAPIA and is promising for automatically assigning keywords to images.




\textbf{Communication Cost}: The communication cost in CPAR comes from two major parts: annotation request and encrypted results returned from the cloud server. The encrypted request consists of a $m_{L1}$-dimensional vector $\textbf{C}(\textbf{V}_{s,L1})$, a $m_{L1}^{'}$-dimensional vector $\textbf{C}(\textbf{H}_{s,L1})$, a ${m}_{KL}$-dimensional vector $\textbf{C}(\textbf{V}_{s,KL})$ and a set of encrypted \textit{split} field elements $\mathcal{SF}$. In the $PCA-32$ setting, the total communication cost for a request is 80KB, in which 26KB for $\textbf{C}(\textbf{V}_s)$, 48KB for $\textbf{C}(\textbf{H}_s)$ and 4KB for $\mathcal{SF}$. Meanwhile, the returned results contain encrypted keywords and distance comparison candidates of top 10 related images. Using AES-256 for keywords encryption, the total size for the returned result is 488 Bytes with the average number of keywords for each pre-annotated image as 5.7. Therefore, the communication cost for each privacy-preserving annotation can be efficiently handled in today's Internet environment.



\section{Related Works}\label{s:related-work}




To solve the problem of how to search over encrypted data, the idea of keyword-based searchable encryption (SE) was first introduced by Song et.al in ref \cite{Song:2000}. Later on, with the widespread use of cloud storage services, the idea of SE received increasing attention from researchers. In ref \cite{EUROCRYPT04,WangCong:2010}, search efficiency enhanced SE schemes are proposed based on novel index constructions. After that, SE schemes with the support of multiple keywords and conjunctive keywords are investigated in ref \cite{Sun-2013AsiaCCS}, and thus making the search more accurate and flexible. Recently, fuzzy keyword is considered in ref \cite{Wang-2014Infocom}, which enables SE schemes to tolerate misspelled keyword during the search process. While these SE schemes offer decent features for keyword-based search, their application to images are limited given the question that how keywords of images can be efficiently extracted with privacy protection. It is impractical for cloud storage users to manually annotate their images. 

To automate the keywords extraction process for images, a number of research works have been proposed with the concept of ``automatic image annotation'' \cite{wang2006annosearch,russell2008labelme,Makadia2010IJCV,eccv12}. Chapelle et al. \cite{chapelle1999support} trained support vector machine (SVM) classifiers to achieve high annotation accuracy where the only available image features are high dimensional histograms. In ref \cite{cusano2003image,shi2004adaptive}, SVM was used to learn regional information as well as helped segmentation and classification process simultaneously. Different from SVM which works by finding a hyperplane to separate vector spaces, Bayesian network accomplishes the annotation tasks by modeling the conditional probabilities from training samples. In ref \cite{vailaya2001image,rui2005novel}, Bayesian networks were built by clustering global image features to calculate the conditional probabilities. Another widely used technique is artificial neural network (ANN). Take ref \cite{park2004content} as an instance, based on the assumption that after image segmentation, the largest part of an image significantly characterizes the entire image, Park et al. annotated images using a 3-layer ANN. With the flourish of deeper ANN structures, such as convolutional neural network (CNN), in various vision tasks \cite{krizhevsky2012imagenet,girshick2014rich,schroff2015facenet}, these deeper frameworks have also been applied to image annotation tasks. In ref \cite{gong2013deep}, Yunchao et al. proposed to solve image annotation problem by training CNN with rankings. Jian et al. \cite{wang2016cnn} combined CNN with recurrent neural network (RNN) to address the problem of the keyword dependency during annotation. However, all of these image annotation works raise privacy issues when delegated to the cloud since unencrypted images need to be outsourced. Therefore, to address such privacy concerns, this paper proposes CPAR, which utilizes the power of cloud computing to perform automatic image annotation for users, while only providing encrypted image information to the cloud.

\section{Conclusion}\label{s:conclusion}
In this paper, we propose CPAR that enables privacy-preserving image annotation using public cloud servers. CPAR uniquely integrates randomized kd-forest with a privacy-preserving design, and thus boosting the annotation efficiency using cloud. Specifically, CPAR proposes the lightweight privacy-preserving $L_1$ distance $PL1C-RF$ and KL-Divergence $PKLC-RF$ comparison schemes, and then utilizes them together with order-preserving encryption to support all required operations in image annotation and randomized kd-forest search. Our $PL1C-RF$, $PKLC-RF$ and privacy-preserving randomized kd-forest can also be utilized as independent tools for other related fields, especially for efficient similarity measurement on encrypted data. Thorough security analysis is provided to show that CPAR is secure in the defined threat model. Extensive numerical analysis as well as prototype implementation over the well-known IAPR TC-12 dataset demonstrate the practical performance of CPAR in terms of efficiency and accuracy.

\bibliographystyle{IEEEtran}
\bibliography{CNS17_Journal}

\begin{IEEEbiography}[{\includegraphics[width=1in,height=1.25in,clip,keepaspectratio]{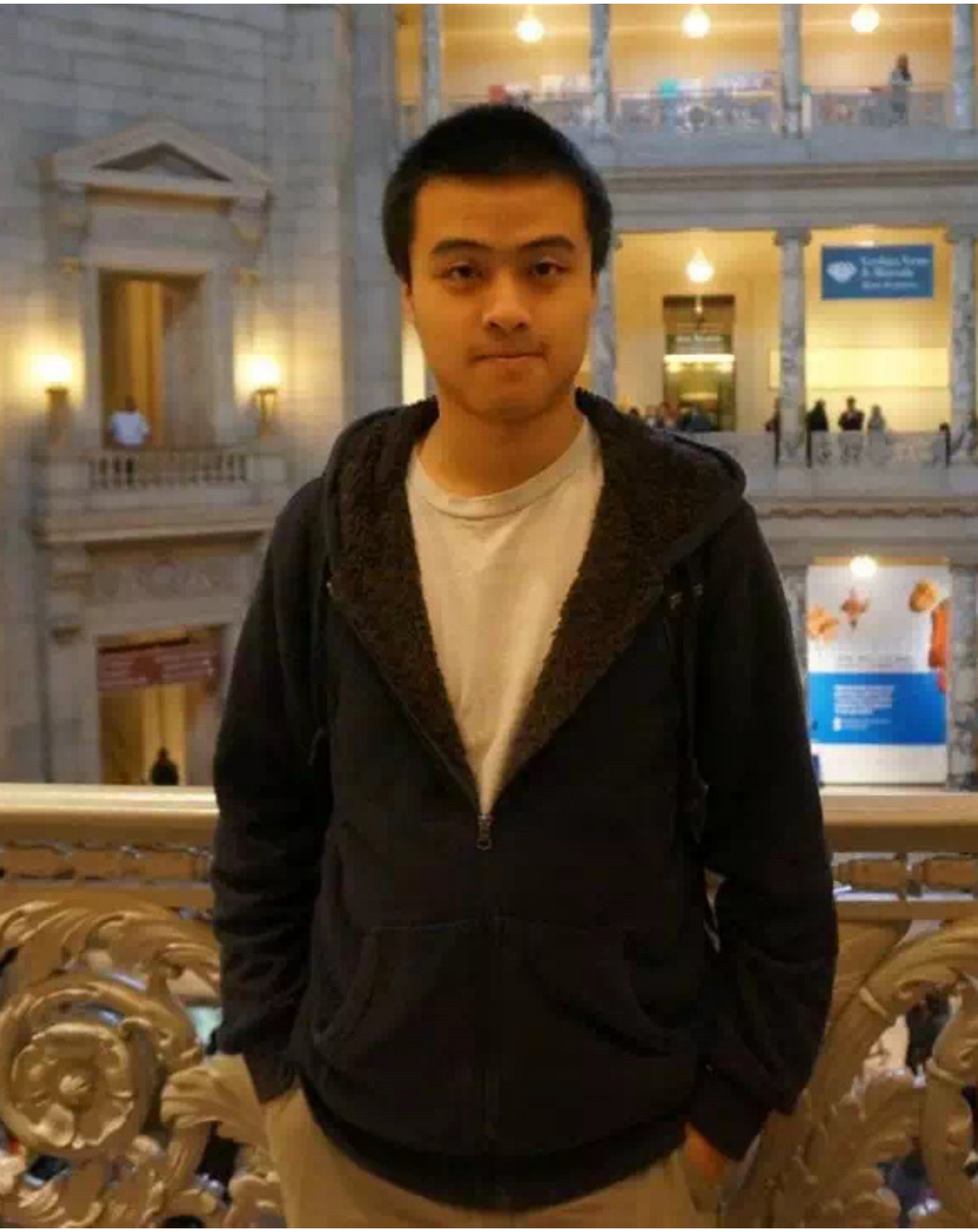}}]{Yifan Tian}
(S'16) is a Ph.D. student at Embry-Riddle Aeronautical University since 2016. He received his M.S. in 2015 from Johns Hopkins University, and B.Eng. in 2014 from Tongji University, China. His research interests are in the areas of cyber-security and network security, with current focus on secure computation outsourcing. He is a student member of IEEE.
\end{IEEEbiography}

\begin{IEEEbiography}[{\includegraphics[width=1in,height=1.25in,clip,keepaspectratio]{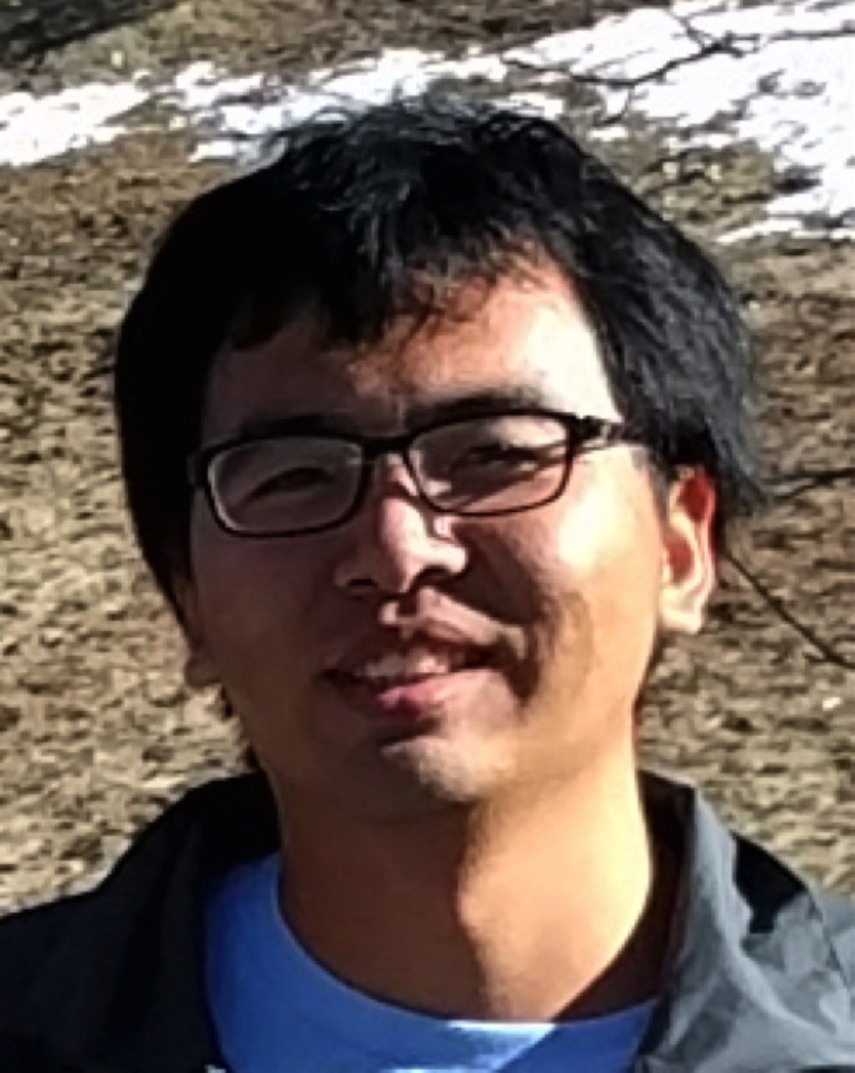}}]{Yantian Hou}
(S'12-M'16) is an assistant professor in Department of Computer Science at Boise State University. He received his B.S. and M.S. degree in Electrical Engineering Department from Beijing University of Aeronautics and Astronautics in 2009 and 2012 respectively. He got his Ph.D. degree in Computer Science Department at Utah State University in 2016. His research interests include wireless network and security, and applied cryptography. He is a member of IEEE.
\end{IEEEbiography}

\begin{IEEEbiography}[{\includegraphics[width=1in,height=1.25in,clip,keepaspectratio]{jiawei.pdf}}]{Jiawei Yuan}
(S'11-M'15) is an assistant professor of Computer Science in the Dept. of ECSSE at Embry-Riddle Aeronautical University since 2015. He received his Ph.D in 2015 from University of Arkansas at Little Rock, and a BS in 2011 from University of Electronic Science and Technology of China. His research interests are in the areas of cyber-security and privacy in cloud computing and edge computing, uav, network security, and applied cryptography. He is a member of IEEE.
\end{IEEEbiography}

\end{document}